\documentclass[superscriptaddress,reprint,aps,prstper]{revtex4-2}

\usepackage{graphicx}
\usepackage{hyperref}
\usepackage{footnote}

\begin{document}

\title{Investigating students’ views of experimental physics in German laboratory classes}

\author{E. Teichmann}
\affiliation{Institut für Physik und Astronomie, Universität Potsdam, 14476 Potsdam, Germany.}
\author{H. J. Lewandowski}
\affiliation {Department of Physics, University of Colorado, Boulder, Colorado 80309, USA}
\affiliation {JILA, National Institute of Standards and Technology and University of Colorado, Boulder, Colorado 80309, USA}
\author{M. Alemani}
\email[]{alemani@uni-potsdam.de}
\affiliation{Institut für Physik und Astronomie, Universität Potsdam, 14476 Potsdam, Germany.}

\begin{abstract}
There is a large variety of goals instructors have for laboratory courses, with different courses focusing on different subsets of goals.  An often implicit, but crucial, goal is to develop students' attitudes, views, and expectations about experimental physics to align with practicing experimental physicists.  
The assessment of laboratory courses upon this one dimension of learning has been intensively studied in US institutions using the Colorado Learning Attitudes Science Survey for Experimental Physics (E-CLASS). 
However, there is no such an instrument available to use in Germany, and the influence of laboratory courses on students’ views about the nature of experimental physics is still unexplored at German-speaking institutions.
Motivated by the lack of an assessment tool to investigate this goal in laboratory courses at German-speaking institutions, we present a translated version of the E-CLASS adapted to the context at German-speaking institutions. We call the German version of the E-CLASS, the GE-CLASS. We describe the translation process and the creation of an automated web-based system for instructors to assess their laboratory courses. 
We also present first results using GE-CLASS obtained at the University of Potsdam. A first comparison between E-CLASS and GE-CLASS results shows clear differences between University of Potsdam and US students' views and beliefs about experimental physics.
\end{abstract}

\maketitle

\section{\label{sec:intro}Introduction}
Laboratory courses are an important part of the physics education curriculum, as they offer the opportunity for students to engage with, and become proficient at, the processes of experimental physics \cite{Trumper03}. 
There is a large variety of teaching goals pursued in laboratory courses. Depending on the institution, one finds courses focused on reinforcing physics concepts, the acquisition of experimental skills, or both of these aspects \cite{AAPT15, AAPT98, Nagel18, landscape}.
In addition to these broad categories of goals, an often implicit, but crucial, goal of laboratory classes, is to have students' views and expectations of experimental physics align with those of practicing experimental physicists. 

Studies of the degree to which students meet these varied goals have been conducted only recently. 
These studies have opened up a discussion about the effectiveness of traditional prescriptive laboratory courses to improve students’ physics content knowledge \cite{Wieman15, Wieman18}, experimental skills \cite{Etk10, Pillay08, Volkwyn08}, and the understanding of the nature of experimental physics \cite{Zwickl13, Zwickl14,WilcoxLew18}. Moreover, it has been shown that having students understand the nature of experimental physics and promoting 'expert-like' attitudes can enhance student motivation and performance in the laboratory \cite{Adams06, Redish98, Perkins05, Lising05}.  
New teaching approaches have been developed, shifting the focus from physics content reinforcement to the acquisition of experimental skills, while engaging students in authentic experimental physics practices \cite{Handelsman04, Etk01, Etk06, Etk07, Holmes15, Zwickl15,Frazer2020}. 
In order to gain insight into students' learning in laboratory classes and evaluate the obtainment of the learning goals, instructors need ways to assess their laboratory classes in an easy and standardized way. The data from these assessments can then be used to guide course-transformations\cite{Zwickl13}. Several research-based assessment tools for laboratory courses have been developed by the physics education research (PER) and are given in Refs. \cite{physport, Day11, Zwickl14, PLIC19, Theyssen16, Theyssen18, Rehfeldt}.
One important example is the Colorado Learning Attitudes Science Survey for Experimental Physics (E-CLASS). It is a research-based survey focused on the nature of experimental physics. It was developed and validated at the University of Colorado Boulder and has been used extensively in the US to assess students’ views and expectations about experimental physics \cite{Zwickl14, Wilcox16,WilcoxLew18}. 

In contrast, one can find only a few research-based surveys for assessing laboratory courses in German-speaking countries \cite{Theyssen16, Theyssen18, Rehfeldt} and none of these tools focus on students' views of the nature of experimental physics.
Some recent studies concern students' views about the nature of science in general, but none of those studies focused on experimental physics \cite{Priemer, Neumann, Woit21}.
The E-CLASS and its current automated administration system \cite{WilcoxZwi16} is not able to be used in German-speaking countries because of the language barrier and the strict European data-protection laws present in the European Union.

Motivated by the lack of assessment tools and studies about students' views about experimental physics in German laboratory courses, we adapted the E-CLASS to the German context. We call the resulting German version of the E-CLASS the GE-CLASS (where ‘G’ stands here for ‘German’).

The main goals of this work are to present the process of this adaptation, the resulting survey, and some initial results from the GE-CLASS from one institution, including a comparison to the US results of E-CLASS. We hope this will motivate German instructors broadly to begin assessing this one dimension of learning in their laboratory classes using the GE-CLASS. Thus, instructors in German-speaking institutions will have data on student learning to help inform their instructional efforts and lab transformations. 
Moreover, by having many instructors use the GE-CLASS, we will gain insight into strengths and weaknesses of laboratory instruction in German-speaking institutions, which could allow for coordinated efforts to improve laboratory instruction.
To this end, we present in this paper (i) how we translated the E-CLASS into German and validated the translation through student interviews, (ii) how we created a centralized survey administrations system for broad dissemination of the GE-CLASS in German-speaking institutions taking into account the constraints of privacy laws in Europe, and (iii) the results from a first exemplary study using the GE-CLASS system at the University of Potsdam.

\section{\label{sec:ECLASS} The E-CLASS}
The E-CLASS consists of 30 core statements, which probe ideas around the nature of experimental physics. Examples include: \textit{“The primary purpose of doing a physics experiment is to confirm previously known results.”}  and \textit{“When I am doing an experiment, I try to make predictions to see if my results are reasonable.”} For each core statement, students answer two types of questions. One is related to their personal view (\textit{“What do YOU think when doing experiments for class?”} referred to as YOU-questions). The other one is related to their perspective of experts’ views (\textit{“What would experimental physicists say about their research?”} referred as EXPERT-questions). Students answer on a five-point likert scale from strongly disagree to strongly agree. Students' answers are compared to the answers given by 23 practicing physicists, referred to here as the EXPERT-reference (ER). Since students complete the E-CLASS before and after the laboratory courses in a pre-/post-test mode, one can study if students’ personal, as well as their perspective of experts, shift upon instruction to more expert-like or not (i.e., if they agree with the ER or not).
An additional set of 23 questions probe students' post-survey perception of the grading practice in the laboratory course. For example, students are asked \textit{“How important for earning a good grade in this class was calculating uncertainties to better understand my results?”}. Instructors can therefore verify if their grading practice is perceived by students in the way they intended.

The E-CLASS probes a wide variety of aspects of experimentation, for example, the importance of measurement uncertainties and systematic errors, self confidence, and the relevance of experiments in physics as a discipline. This large variety of aspects makes the E-CLASS applicable in diverse instructional laboratory environments with different underlying goals, as instructors are provided with aggregate student responses to each individual statement and are encouraged to focus on only the statements that align with their particular goals for the class. 
Many research studies have been performed using the E-CLASS within recent years \cite{Zwickl14, WilcoxLew16, WilcoxLew17, WilcoxZwi16, WilcoxLew18, MichaelJFox}, with 95\% of the E-CLASS dataset coming from US institutions. The entire deidentified data set has been recently made publicly available \cite{Aiken&Lew}.  
Some of the main results of those studies are the following:
\begin{itemize}
    \item Even when students have novice personal views about experimental physics, they know how an expert would respond to the statements \cite{Zwickl14}.
    \item In traditional, guided, first-year lab courses, a decrease in the mean overall E-CLASS score from the pre-test to the post-test is observed \cite{WilcoxLew18}.
    \item Courses that aim to primarily improve experimental skills score higher on the E-CLASS than courses that aim to primarily reinforce concepts \cite{WilcoxLew17}.
    \item Partially open-ended labs show higher E-CLASS scores than fully guided lab courses \cite{WilcoxLew16}.
    \item Perception of the grading scheme is correlated with E-CLASS performance \cite{WilcoxLew17PERC}.
\end{itemize}

A more complete review of the results from a large selection of studies done with the E-CLASS can be found in Ref. \cite{WilcoxLew18}.

\section{\label{sec:background}Translation of the E-CLASS to German}
In order to adapt the E-CLASS to the German-speaking context, we performed a multi-step iterative translation accompanied by think-aloud student interviews. The process started by having two German-native-speaking experimental physicists independently translate the E-CLASS into German. They then discussed the discrepancies until they reached 100\% consensus. This German translation was translated back to English by an English-native-speaking physicist. These three translators are not authors of this paper.  The back translation was then compared to the original E-CLASS. The discrepancies between the two versions were analyzed and discussed by the translation team and one of the authors (M.A.) until a consensus was achieved. The resulting German translation was validated through think-aloud interviews with students. A total of 12 think-aloud interviews with physics major students were conducted by M. A.. Among the students interviewed, four were women, eight were men; six were first-year students and six were second-year students. All of the students volunteered to take part in the interview process by responding to an announcement after laboratory classes at the University of Potsdam.
In the interviews, students were asked to read in order the items of the survey and for each item explain their understanding of that item. We did not interrupt the students unless it was not completely clear to us what students meant. In this case, we asked for clarification or to give an example of what they meant.
The interviews were audio-recorded for the data analysis. Authors M.A. and H.L. (original E-CLASS survey developer) discussed the results of the translations and interviews to assure that students' interpretations of the survey items had the same meaning as the original English version. In case of discrepancies between the German and English versions, we refined the translation together with the translators before further interviews were performed.
We note that H.L. is a native English speaker with only a basic level knowledge of German. M.A. is not a native German speaker, but is fluent, and has taught lab courses, in both German and English. After the last set of changes, students were interpreting the statements consistently and as we intended.
A complete list of the translated items on GE-CLASS can be found in Appendix A.

In the following, we report some examples of the minor changes we made during the entire iterative translation process to obtain the final German survey version. 
Those changes arose for different types of reasons. In some cases, we found mistakes or not literal translations. For example, for Question 8 \textit{'When doing an experiment, I try to understand the relevant equations'} the word 'equations' was first incorrectly translated as 'questions' and for Question 4 \textit{'If I am communicating results from an experiment, my main goal is to create a report with the correct sections and formatting'} the word 'sections' was first translated as 'outline' (German \textit{'Gliederung'}) instead of 'sections' (German \textit{'Abschnitten'}). For some items, the two German translators used very slightly different translation options. In such cases, we decided on one of the two versions, but we kept the other version ready during the interviews in case students encountered difficulties with the chosen version. 
A third reason for changes was when students had comprehension problems.
For the majority of the survey items, students were able to interpret as intended the German version during the interviews. However, difficulties arose with items Q14, Q15, Q22, Q24 and Q26. Below, we explain the issues with those items and any resulting change:

\textbf{Q14:} (\textit{‘When doing an experiment I usually think up my own questions to investigate’.})
In the first interviews, this question was misinterpreted and the word 'questions' was interpreted as 'a comprehension question students have when they read the lab manuals'. We had to change the wording to clarify (by using the word 'erforschen') that the 'questions' are open (research) questions one can answer by doing an experiment.

\textbf{Q15:} (\textit{‘Designing and building things is an important part of doing physics experiments'.}) 
Students were sometimes interpreting our translation as meaning 'essential part' instead of 'important part'. We changed the translation of 'important' from 'wesentlicher' into to 'wichtiger' to address this slight difference in meaning.

\textbf{Q22:} (\textit{‘If I am communicating results from an experiment, my main goal is to make conclusions based on my data using scientific reasoning'.}) This item was often hard to understand for students initially without extra time for reflection.
To help students understand this item easily, we created three slightly different versions of the statement. We asked students (starting from the second interview) each one in turn and had them discuss which one was easiest for them to understand.  We kept the version with the highest rating (we had 10 students out of 11 choosing the final version).

\textbf{Q24:} (\textit{‘Nearly  all  students  are  capable  of  doing a physics experiment if they work at it.’}) 
Two interviewees out of 12 were unsure if this statement was meant to check if the lab manuals/equipment in the laboratory were good enough so that students could succeed in the experiment or (as actually intended by the item) if the statement was referring to students' skills in experimentation. The other 10 students understood this sentence immediately as intended as if students have the skills to perform physics experiments. We kept the translation as it was, as the other meaning understood by those two students cannot be ruled out even in the English version of this item and both students mentioned that they were unsure about the meaning of the item but reported both possibilities.

\textbf{Q26:} (\textit{‘It is helpful to understand the assumptions that go into making predictions.’}) similarly to Q22, many interviewees needed to think longer on this item, but all understood it as intended. The difficulty in the comprehension was due to the fact that students had to think about the meaning of the words 'assumption' and 'prediction'. To facilitate the understanding of the word 'assumption' (which was translated with 'Annahmen') we added 
the adjective 'assumed' ('vorausgesetzte') to it.
The rest of the items were easily interpreted as intended by all students interviewed and were not changed for the final version shown in Appendix A.

\section{\label{sec:}Centralized Automated  system}
We have created an online, centrally administrated and automated system for instructors to use the GE-CLASS, with the goal of encouraging and facilitating its use in German-speaking institutions.
Centralized, automated assessments systems have two main advantages over paper and pencil assessments \cite{WilcoxZwi16,vandusen}. First, automated systems reduce the effort required of instructors using the assessments in their courses by collecting the data and performing the analysis. Second, centrally administered systems can provide additional information to instructors, i.e., the comparison between their course results and the results from similar courses at other institutions\cite{WilcoxLew18}.
With this in mind, we have created a web-page front-end for instructors to register their courses with a back-end in which an automated data analysis system is integrated. The link to the webpage can be found in Ref.\cite{WebpageGE-CLASS}.
The web server, as well as the analysis for the report, are programmed in Python3~\cite{10.5555/1593511} using flask (for the web server) and the SciPy~\cite{2020SciPy-NMeth,2020NumPy-Array,4160265} stack for the analysis.
When setting up a new course, instructors are asked to answer a series of questions about their classes and instruction methods. We collect information about the course, such as type of institution, how many weeks students work in the lab, how many different experiments they perform, how many projects they do, goals of laboratory classes, and what kind of activities students do in the laboratory course.

Once a course is registered, the automated system assigns an identification code to the course that instructors have to provide to their students together with the web-link for the student survey. 
It is important to note here that the way we constructed the GE-CLASS system was discussed beforehand with the ethics commission and the data-protection-office of the University of Potsdam in order to respect the ethic codes of the University of Potsdam and the European Union General Data Protection Regulation (EU-GDPR).
One result of this process was the obligation to assure student anonymity. When taking the survey, students have to create a self-generated anonymous code (composed by 3 letters and 2 numbers) by answering to 3 questions:
\begin{enumerate}
    \item Last two letters of the name of your father (Hara\textbf{ld})
    
    \item The second letter of the place in which you were born (B\textbf{e}rlin)

    \item The day of your mother’s birthday (\textbf{26}.12.1978)

\end{enumerate}
The code for the example above would be \textbf{lde26}. 

For each student, we assign two identification codes. One ID of the course and one (anonymous) for the person. 

We often found that students codes in the pre- and post-survey differed by 1 number or letter, suggesting that some students produced an erroneous code either in the pre- or post-survey. Those data are unfortunately lost because they cannot be matched and used for the data analysis (see section \ref{sec:data analysis} below). We are now thinking about ways to avoid this problem, either by changing the questions to generate the code or suggesting students to write their code on their labbooks to remember. We also note that the questions asked above have to be considered in a cultural context of Germany. We acknowledge that these questions would not necessarily be appropriate in a US context.

Two weeks after the post-test start-date, instructors can log on the GE-CLASS web-page and find the analysis of their course data in form of a report.
In the report, an overview of the results of the course is provided at the beginning. For example, one can find in the overview how many students did the pre- and post- survey, how many students' surveys were considered valid for the data analysis (see section \ref{sec:data analysis} below) and how many surveys were used to provide a comparison with similar courses. The first graph of the report shows the overall agreement with experts averaged over all items and all students. Note that in the report, we do not provide a comparison with US results, but for each graph we provide the comparison with the data set of similar courses that we collected in the GE-CLASS system.
In the second and third graph, one can see the agreement with experts for each item for the YOU-questions and EXPERT-questions respectively. 
In the fourth graph, we report for each item the comparison between results of the YOU- and EXPERT-questions.
In the last graph, the results about students' grading perception for each survey item are shown.
An example report can be found at \cite{report}.

\section{\label{sec:First results}First study using the GE-CLASS system}
Using the above described GE-CLASS system, we have obtained the first results studying the physics laboratory classes at the University of Potsdam. The data have been collected for three consecutive semesters: Winter 2019, Summer 2020, and Winter 2021. 

\subsection {Course Contexts}
We collected data from nine different instances of laboratory courses (see table \ref{tab:table1}). All of the instances of courses are part of the introductory physics laboratory course sequence for physics major students, either studying in the bachelor of science program (B.Sc.) or in the bachelor of education program (B.Ed.). Physics students at University of Potsdam must take four courses belonging to the introductory lab course sequence during the first four semesters, therefore during their first and second year of the university. We have indicated in the Tab. \ref{tab:table1} which course corresponds to which semester. For each semester course, we have a set of different learning goals and have structured the courses differently. Note that in each semester, courses for physics students in the B.Sc. and B.Ed. have the same goals and content, but B.Ed. students have fewer weeks of labs. We refer to courses with the same goals and structure being a ‘type of lab course’. Details about the learning goals and structure of each ‘type of lab course’ can be found below.  We have also reported in Tab. \ref{tab:table1} in which modality the courses were conducted (online, in-person, or hybrid) and have indicated with a asterisk if the courses were affected by the COVID-19 pandemic. The main aspect that changed during the pandemic was that students conducted the experiments alone in the lab and/or at home (when the restrictions occurred) and not in groups as usual. The course Type C was created during the pandemic as a solution for letting students conduct the experiments at home, and because of the positive student response, we kept this type of course in our program.
In the last two columns of Tab.\ref{tab:table1}, we show for each course (i) how many matched pre-/post-test student responses we collected for the data analysis and (ii) what percentage of the class this number correspond to. 
Due to the data privacy rules, we are not able to give course credit for completion of the survey, which would increase participation rate as it was shown for the E-CLASS \cite{Zwickl14}. This fact explains, in part, the low percentages in the last column. Another cause for the low matching rate is, as stated below, that students ID codes often do not match. To have higher participation rate, we recommend allocating extra time during class to complete the surveys. In fact, courses in which we gave in-class time (indicated with a symbol **) show higher participation rates.

\begin{table*}[ht]
\caption{\label{tab:table1}
Description of the the nine instances of the courses from which data have been obtained. We distinguish between courses offered for students in the first year (FY) and beyond the first year (BFY). We indicated the semester in which students take the courses, the study program (Bachelor of Science (B.Sc.) and Bachelor of Education (B.Ed.)), and course goals and structure (referred as course type). A detailed description of the course types can be found in the text. In the table, we also show, for each course, how many weeks the lab course lasted, the course modality (in-person, online, hybrid) and give details about how many GE-CLASS valid responses have been obtained for each course and the response rate.
}
\begin{ruledtabular}

\begin{tabular}{ccccccccc}

&Year&Semester&Course Type&Study Program&Weeks&Modality &Matched Answers  &Participation Rate \\

\hline

\colrule
Course 1 & FY & 1st. & A & B.Sc. & 6 & In-person & 37 & 74\% **\\
Course 2 & FY & 1st. & A & B.Sc. & 6 & Hybrid* & 25 & 48\% ** \\
Course 3 & FY & 1st. & A & B.Ed. & 4 & Online* & 21 & 51\% ** \\
Course 5 & FY & 1st. & A & B.Ed. & 4 & In-person & 5 & 33\% ** \\
Course 9 & FY & 2nd. & B & B.Sc. & 9 & Hybrid* & 1 & 3\% \\
Course 4 & BFY & 3th. & B & B.Sc. & 7 & In-person* & 12 & 30\% \\
Course 8 & BFY & 3th. & B & B.Ed. & 4 & In-person* & 2 & 20\%\\
Course 7 & BFY & 4th. & C & B.Sc. & 8 & Hybrid* & 3 & 14\%\\
Course 6 & BFY & 4th. & C & B.Ed. & 6 & Online* & 5 & 33\% \\
\end{tabular}
\end{ruledtabular}
\end{table*}

Here, we describe briefly the general learning goals of our laboratory course, as well as the specific learning goals of each of the four parts of our introductory laboratory course curriculum and the course structure. 

\subsubsection {Description of type of courses}
In 2016, we started a process of restructuring our laboratory courses at University of Potsdam, focusing on teaching students experimental skills and offering an authentic experience of the processes of experimental physics.  Examples of these experimental skills are modeling and design, as well as ‘critical thinking’ and ‘problem solving.’ Furthermore, we want students to understand the nature of experimental physics and learn to ‘think like a scientist.’ 
(Note that the reconstruction of our courses is a 'work in progress' and some experiments offered to students in course Type B have not been transformed yet. We are using the results presented of this work, to guide further our course transformation.)
Here, we describe briefly the specific learning goals and structures of the different types of courses investigated in this study.

{\bf Course Type A:} 
We had three primary specific goals for the first semester courses. We wanted students to:
\begin{enumerate}
    \item be able to apply the basic concepts of measurement uncertainties and systematic errors.
    
    \item to be able to write a good lab notebook and explain why physicists use lab-books.

    \item be able to recognize what components make up a good graph, be able to create a good graph, and be able to use graphical analysis methods (for example, least squares fitting and linearization).

\end{enumerate}

This type of course included both a seminar component with exercises, group work, quizzes, peer-instruction, and homework problems and a laboratory component. In the lab, students performed experiments that had them practice working with measurement uncertainties, systematic errors, graphing, and writing lab-books. The experiments often had answers or end results that were initially unknown to the students. For example, students had to measure and compare the reaction time for visual stimuli of their group members. They were also asked to measure their reaction time using another measurement method, compare the results of the two methods (one method having an unknown systematic error), and discuss which of the two methods was more precise and which method was more accurate. In this setting, the lab guides had supporting questions to guide students' work.

 {\bf Course Type B:} 
In the second and third semester courses, we want students to understand the role of modeling in experimental physics and practice it during their lab work\cite{Dounas_Frazer_2018}. To engage students in the process of modeling, we made use of the ‘Modeling Framework for Experimental Physics’ described in the literature \cite{Zwickl15,Dounas_Frazer_2018}. 

In the second semester, in addition to laboratory work, we had a seminar session in which we introduced the Modeling Framework for Experimental Physics \cite{Zwickl15}. In the third semester course, we used only the laboratory setting. 

As an example of a lab for this type of course (two three-hour lab sessions), students had to design, build, and calibrate an optical spectrometer capable of resolving the sodium doublet. For this lab, students were provided with an optical table, several lenses and gratings, a line camera, and atomic lamps like Na, Ne and Hg.

 {\bf Course Type C:} 
In this type of course, we want students to continue engaging with modeling, but also work on experimental design and communication. The additional goals include:
\begin{enumerate}
    \item Learn how to use and program microcontrollers (Arduino) and sensors to perform simple experiments with them using the process of modeling. 
    
    \item Practice how to plan and design an experiment based on students' own ideas and write a proposal about it.

\end{enumerate}
After an introduction session on Arduinos microcontrollers and their programming language, students performed several experiments (at home) that aimed to foster modeling. Students were given little guidance on how to set up experiments in their home and how to take the data, but were given a specific problem to work on and were encouraged to engage in the process of modeling. In the second part of the course, there was one session focused on how to write proposals. Students then planned their own experiment using Arduinos and sensors, wrote a proposal about it, and peer-reviewed other students' proposals. For course 7, students were given time for the realisation of their idea in the lab.

\subsection{\label{sec:sample}Data sets}

\subsubsection{\label{sec:samplegeclass}GE-CLASS}

Using the GE-CLASS system, we obtained 111 valid student survey responses. Responses are considered valid only in case the anonymous code and the course identification code described above matched in the pre-/post-tests. Additionally, a control question was used to validate pre- and post- tests independently, in which students are explicitly asked to respond with a particular answer. Among the 111 responses, 89 responses are from courses for the first year (FY) and 22 are from courses beyond the first year (BFY) (Tab \ref{tab:table1}).

\subsubsection{\label{sec:sampleeclass}E-CLASS}
 One of our research questions is to compare the sample of responses at the University of Potsdam to responses from US institutions broadly. Since we do not have enough data to be able to split our GE-CLASS data into FY and BYF subsets, and all of the students at Potsdam considered in this study are physics majors, we select data from the US dataset to make more direct comparisons. To do this, we first filter the US data to include only physics majors. Then, we randomly choose students in both FY and BFY courses such that the ratio of these two populations is the same for the US dataset and Potsdam dataset. We note this is a limitation of our work in Sec.\ref{sec:limiations}.

The details of the resulting filtered datasets for the E-CLASS and GE-CLASS are shown in Table \ref{tab:table2}.

\begin{table}[h!]
\caption{\label{tab:table2}%
Data used in this study. The numbers for GE-CLASS represent all of the data from University of Potsdam. The data from E-CLASS represent a filtered dataset to match the ratio of FY to BFY responses present in the GE-CLASS data set. All data are from physics majors.
}
\begin{ruledtabular}
\begin{tabular}{ccc}
&GE-CLASS&E-CLASS\\
\hline
Institutions & 1 & 90\\
FY Courses & 4 & 306\\
BFY Courses & 5 & 241\\
Matched student responses (FY) & 89 & 1073\\
Matched student responses (BFY) & 22 & 264\\
Total students responses & 111 & 1337\\
\end{tabular}
\end{ruledtabular}
\end{table}

\subsection{\label{sec:data analysis}Data Analysis}
We used two different schemes for comparing students' answers to the ER. One for creating the reports for instructors (i.e., for instructional purposes) and one for investigating our results for research purposes. We present both types of analysis as the stated goals for this paper include encouraging instructors to use GE-CLASS, so it is important to present results that they will see in their reports, and beginning to investigate the impact of courses on students' views for use by education researchers, so we include the second format of our results. 
In both cases, we first reduced the five-point Likert scale (see section \ref{sec:ECLASS}) to  a three-point Likert scale, by collapsing 'strongly (dis)agree' and'(dis)agree' into a single category.
For the instructor reports, we used a binary scale analysis scheme. This was realized by assigning a numerical score of +1 for answers in agreement with the ER and a score of 0 otherwise. Such a binary analysis is easy to interpret for instructors as it directly provides information on what percentage of the class agrees with experts and has an easy to interpret visual representation. For research studies, we use a three-point scale analysis. Here, we assign a numerical score of +1 for agreement with the ER, 0 for neutral, and -1 for disagreement with the ER. These schemes are the same ones used for instructors and researchers for E-CLASS.

In this paper, the results in Figs. \ref{fig1} and \ref{fig2} are obtained using a two-point analysis, while results in Figs. \ref{fig3} and \ref{fig4} are obtained using a three-point analysis.

In this paper, we used two statistical (non parametric) tests, the Mann-Whitney U-test \cite{MannWhitney47}, and the Anderson-Darling test for k-samples \cite{K-SampleAnderson}.
The Mann-Whitney U procedure was used to test if the observed differences between two sample distributions were statistically significant.
We used the null hypothesis that the two samples are coming from the same population. We rejected the null hypothesis for a p-value p$<$0.05. To evaluate practical significance, we calculate an effect size using Cohen's d\cite{Cohen88}. 
In order to distinguish if the cumulative distributions in Fig. \ref{fig4} were statistically different, we used the Anderson-Darling k-samples test, with null hypothesis that the two-samples are drawn from the same population. For this test we used k=2 and a significance level of 0.05.

\subsection{\label{sec:results}Results}

\begin{figure}[b]
\includegraphics{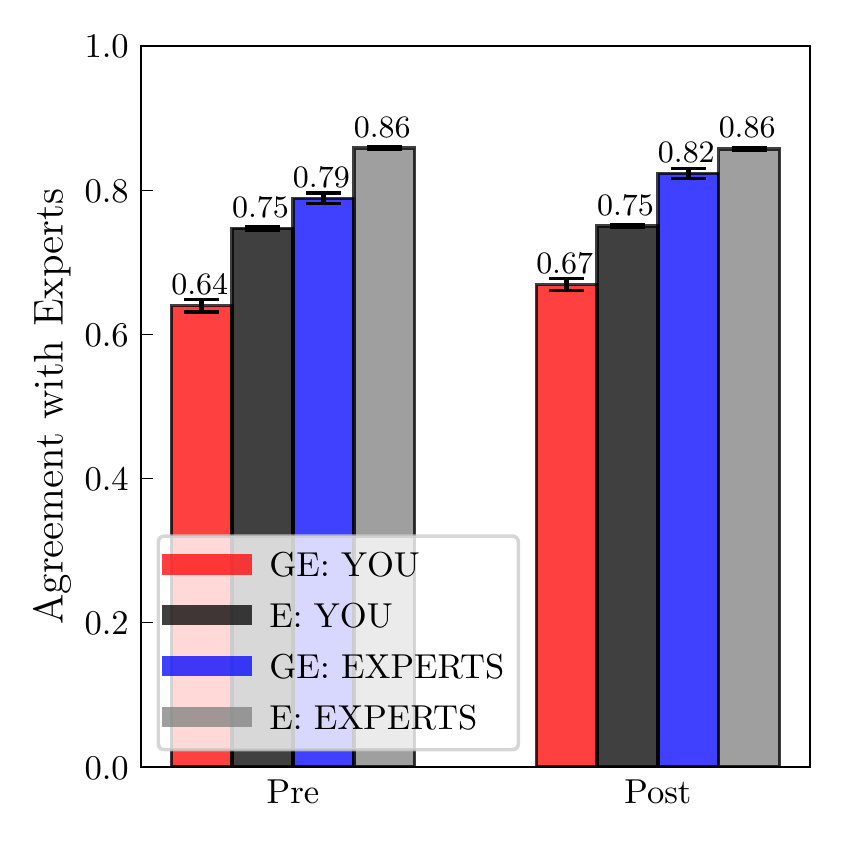}
\caption{\label{fig1} Overall GE-CLASS agreement with experts for the pre- and post-tests. Perfect agreement with experts corresponds to one, zero means no agreement with experts. In red are the GE-Class results for YOU-questions and in blue for the EXPERT-questions. For comparison, E-CLASS results are indicated in the figure as well (in dark and light grey). The overall mean shown here averages over all students and all items on the survey. Error bars are standard errors.}
\end{figure}

The overall agreement with the ER is shown in Fig. \ref{fig1}.
It was calculated using the 2-point scale analysis described above, by averaging over all students and all items. 
Figure \ref{fig1} shows both results for students’ personal views, as well as students’ perspective of experts’ (labeled as ‘YOU’ and ‘EXPERT’ respectively) for both the GE-CLASS and E-CLASS (labeled as ‘GE’ and ‘E’ respectively). 
This type of graph is also provided to instructors by the automated GE-CLASS system \cite{report}.

For the YOU-questions, the overall GE-CLASS agreement with the ER is 0.640$\pm$0.008 before instruction and 0.667$\pm$0.008 after instruction (see Fig.\ref{fig1}). This shift is statistically significant, as tested using a Mann-Whitney U-test (p=0.0058$\ll$0.05)  \cite{MannWhitney47} with a very small effect size (Cohen’s d=-0.062 \cite{Cohen88}).

We also observe that students’ perspectives of what experts think (EXPERT-questions) have much larger pre-instruction agreement with the ER (0.789$\pm$0.007) than the YOU-questions. The students' perspective of experts is also positively influenced by instruction, as we  measure 0.823$\pm$0.007 for the post-test. This positive shift is statistically significant with a Mann-Whitney U-test p$\ll$0.05 and a very small effect size (Cohen’s d=-0.087).

In Fig. \ref{fig1}, we also show the results from the E-CLASS data set.
We found that the overall E-CLASS agreement with ER for the YOU-questions both before and after instruction are higher than the scores from the University of Potsdam. Before instruction, the E-CLASS overall score is 0.747$\pm$0.002. The difference between E-/GE-CLASS is statistically significant (Mann-Whitney U-test p$\ll$0.001) with a small effect size (Cohen's d=-0.24).
After instruction, the E-CLASS overall score is 0.750$\pm$0.002.
This difference between E-/GE-CLASS results is also statistically significant (Mann-Whitney U-test p$\ll$0.001) with a small effect size (Cohen's d=-0.20).

For the EXPERT-questions, the overall expert agreement score before instruction is for the E-CLASS 0.858$\pm$0.002. This value is higher than for the GE-CLASS. The difference between these values is statistically significant (Mann-Whitney U-test p$\ll$0.001) with a small effect size (Cohen's d=-0.20). However, the gap between E-/GE-CLASS results becomes smaller for the EXPERT-questions after instruction (0.857$\pm$0.002 for the E-CLASS). The difference is statistically significant with a smaller size effect than for the pre-instruction (Cohen's d=-0.10).

It is important to note, that no change upon instruction (for both YOU- and EXPERT-questions) is observed in the E-CLASS for the data set used here.

\begin{figure*}
\includegraphics{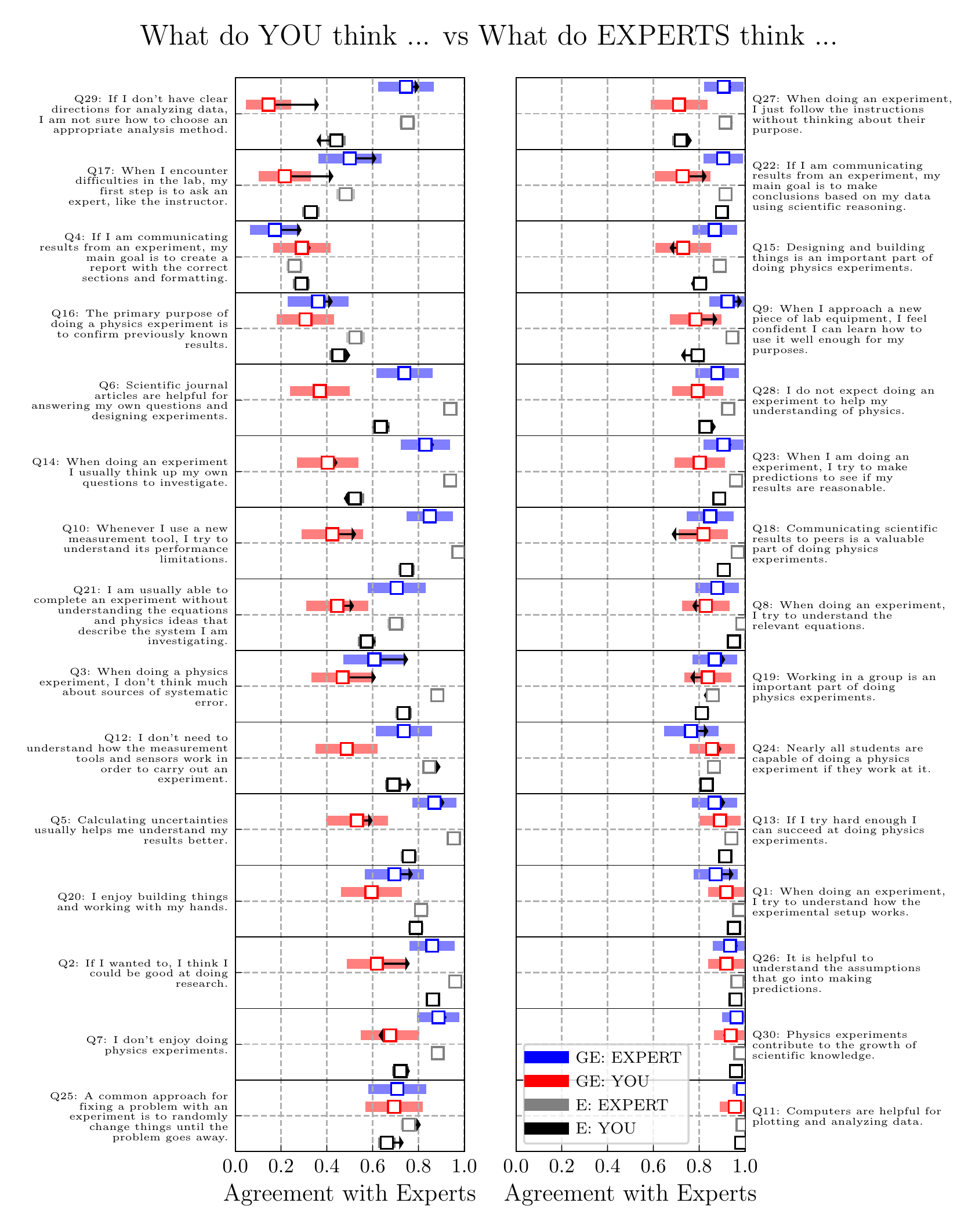}
\caption{\label{fig2} Fraction of responses that agree with the ER for each item of the survey. Squares indicate the mean for the pre-instruction responses for YOU-question (blue) and EXPERT-question (red) views. Black and grey circles are the E-CLASS pre-instruction responses for YOU-question and EXPERT-question view respectively. Shaded bars indicate the 95\% confidence intervals of the pre-instruction fraction. These intervals are not visible for E-CLASS results since they are smaller than the data point symbols. The end of the arrows indicate the post-instruction agreement with experts. Note, the questions in the figure  are ordered according to the pre-instruction personal views for the GE-CLASS data.}
\end{figure*}

\begin{figure*}
\includegraphics{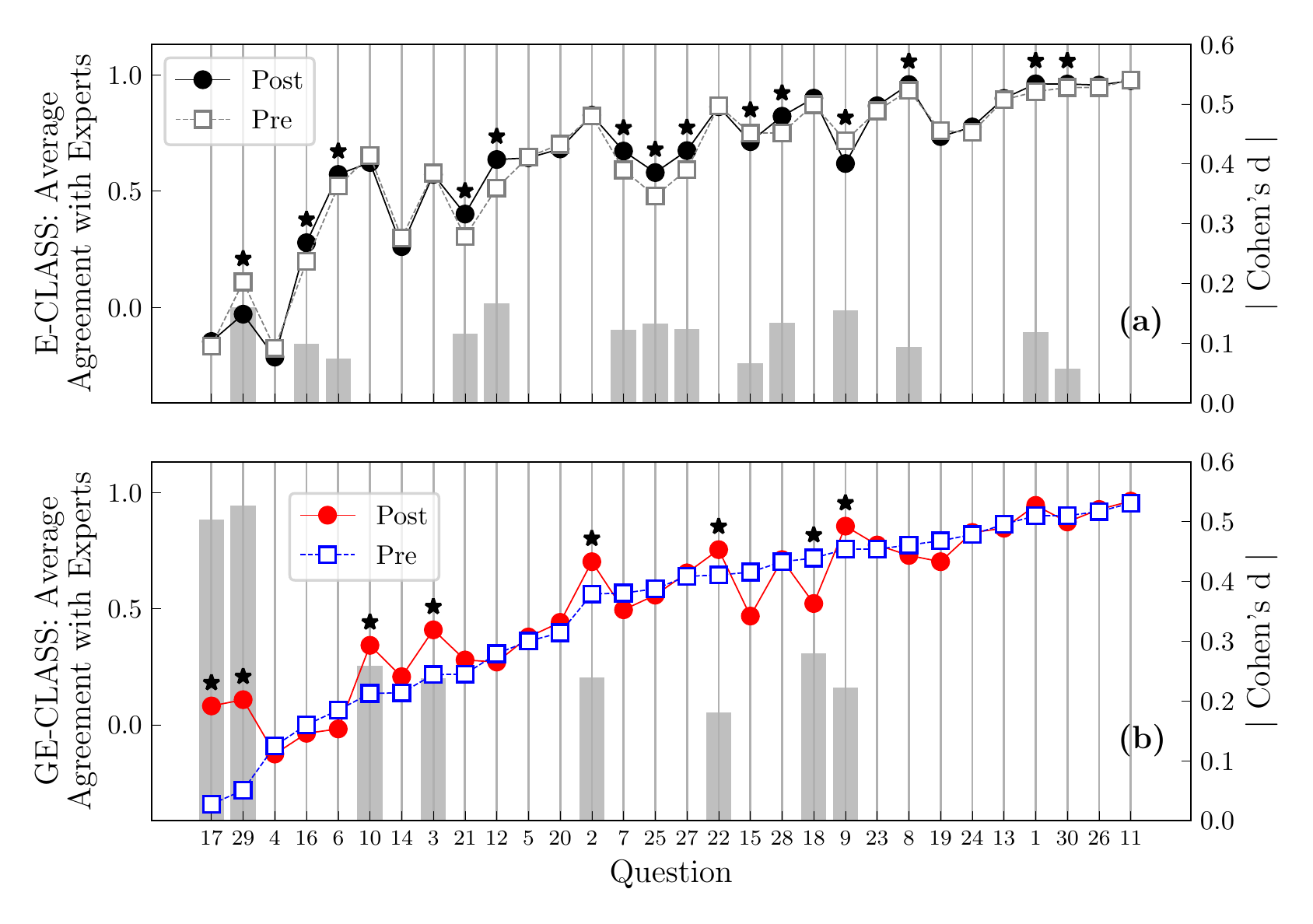}
\caption{\label{fig3}Overall agreement with the ER for YOU-questions, item by item, before and after instruction for E-CLASS (top graph) and for the GE-CLASS (lower graph). The results are obtained using a 3-point scoring method (left axis). Stars indicate pre-post changes that are statistically significant, as obtained by using the Mann-Whitney U-test. The absolute values of the Cohen’s d  are also indicated in the figure as bars for the items with a statically significant change after instruction (right axis).}
\end{figure*}

\begin{figure*}
\includegraphics{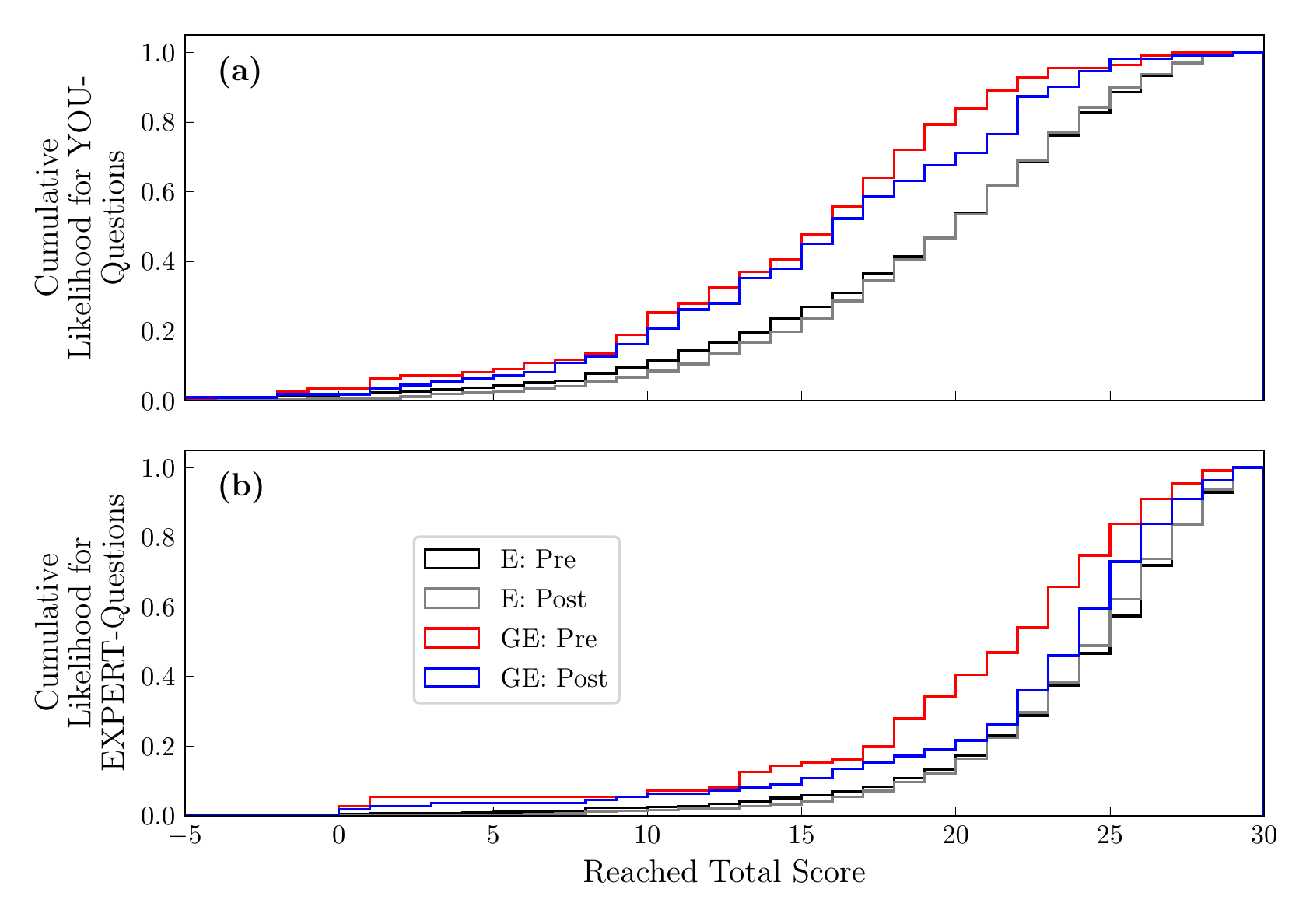}
\caption{\label{fig4} Cumulative distribution of the agreement with experts as a function of total GE-/E-CLASS score for pre-/post-tests in (a) for YOU-questions and (b) for EXPERT-questions.  The reached total score has been calculated using a 3-point data analysis (see text for more information).}
\end{figure*}

In Fig. \ref{fig2}, we present a detailed item-by-item analysis of the GE-CLASS responses. 
This type of graph is also provided to instructors by the automated GE-CLASS system \cite{report}. Once again, here, a 2-point scale analysis was used.
Results are ordered based on the level of agreement with the ER in the pre-instruction YOU-questions. In figure \ref{fig2}, we also show the results of the E-CLASS for comparison.

The EXPERT-questions show larger agreement with the ER than the YOU-questions in general and, in particular, for questions Q29, Q6, Q14, Q10 and Q5 in both E-CLASS and GE-CLASS data. Thus, there is a pronounced gap between students personal view and students' knowledge of experts' views for these questions.

Moreover, we observe that in 30\% of the survey's items, the values of ER-agreement of YOU-questions in the GE-CLASS data are significantly lower than those of the E-CLASS results. For the EXPERT-questions, this is the case for about 7\% of the items. For determining significance here,  we considered the 95\% confidence intervals indicated in Fig.\ref{fig2}. Therefore, we found that students personal beliefs at the University of Potsdam are, for a considerable amount of items of the survey, lower than for students participating in the E-CLASS.

Looking now at changes upon instruction, we observe the largest positive shifts in the GE-CLASS for the YOU-questions Q29 (\textit{‘If I don't have clear directions for analyzing data, I am not sure how to choose an appropriate analysis method’}) and Q17 (\textit{‘When I encounter difficulties in the lab, my first step is to ask an expert, like the instructor’}.
Those items are followed by Q3 (\textit{‘When doing a physics experiment, I don’t think much about sources of systematic error.’}) and Q2 (\textit{‘If I wanted to, I think I could be good at doing research’}).
The largest negative shift is observed for question Q18 (\textit{‘Communicating scientific results to peers is a valuable part of doing physics experiments’})

To gain further insights about the item-by-item E-/GE-CLASS comparison, we looked at the differences between E-/GE-CLASS results using a 3-point scale analysis as described above. The results are shown in Fig \ref{fig3}. The survey’s questions in the figure have been ordered based on the values of the pre-instruction GE-CLASS average agreement with the ER.
Statistically significant shifts upon instruction are indicated by stars in Fig. \ref{fig3}. The effect size of those statistically significant shifts have been analysed by calculating the Cohen's d values and are indicated in the figure using grey bars referring to the right axis. For the pre-instructions data in Fig.\ref{fig3}, the order of items from low to high agreement with ER show a similar trend for E-/GE-CLASS results. Interestingly, the four questions with the lowest agreement, as well as the three questions with highest agreement, are the same questions for E-/GE-CLASS. This indicates that both groups score highest and lowest (i.e., have expert-like/non-expert-like thinking) on items regarding the same aspects of experimental physics. Additionally, we found that for seven items there are statistically significant positive shifts (see Fig. \ref{fig3}), while for one item (Q18 (\textit{‘Communicating scientific results to peers is a valuable part of doing physics experiments’})) there is a statistically significant negative shift. 

For the E-CLASS data in Fig. \ref{fig3}, a statistically significant positive shift occurs for 11 items, while for three of the items a statistically significant negative shift is observed (for items 29, 15 and 9). For those items, slightly larger effect sizes are found for the GE-CLASS than for the E-CLASS results, as indicated by the grey bars in Fig. \ref{fig3}, which show the values of the Cohen’s d.

To further compare the E-/GE-CLASS results and the changes upon instruction, we analysed the changes of the cumulative likelihood distribution of the total score. This allow us to examine the distribution of scores and not just the mean. This distribution is shown in Fig.\ref{fig4} for both before and after instruction for both the YOU- and EXPERT-questions. This type of graph shows the integrated fraction of students that received up to a certain total score on the survey, where a student's reached total GE/E-CLASS score is obtained by summing up the scores on each of the 30 items using a 3-point analysis scheme. The reached total score can therefore range from a minimum of -30 (disagreement with experts for all survey items) to a maximum of +30 (agreement with experts for all survey items). The best possible distribution would be zero everywhere and a sharp peak at ‘Reached Total Score’= +30. This would mean that all students agree for all items with experts. Additionally, the smaller the area under the curve the more the distribution of responses that aligns with experts.

Comparing the E-/GE-CLASS results before instruction, we see that the point where the cumulative distribution of students at University of Potsdam begins to increase is shifted to the left with respect to the one for the E-CLASS data, meaning that more students at University of Potsdam start with lower survey scores. 
Moreover, the curve reaches near 1.0  at a lower Reached Total Score for the GE-CLASS data, meaning that the highest survey scores in the GE-CLASS dataset are lower than for the E-CLASS dataset. This trend is particularly pronounced for the YOU-questions. At University of Potsdam, fewer students reach the highest survey scores before instruction (the maximum total reached score is in fact for the YOU-questions +28) as compared to the E-CLASS results where there are students who reach up to +30.

Using an Anderson-Darling k-samples statistical test, we investigated if the observed differences between the E-CLASS and GE-CLASS cumulative distributions in both pre- and post-surveys and for YOU- and EXPERT-questions are statistically significant. We found for all these cases that the two samples of the E-/GE-CLASS do not originate from the same distribution. For both YOU- and EXPERT-questions, we observe larger shifts upon instruction of the distribution for  GE-CLASS as compared to the E-CLASS. Anderson-Darling k-samples statistical test indicate that changes upon instruction are statistically significant only in the case of the EXPERT-questions of the GE-CLASS data.

\subsection{\label{sec:discussion}Discussion}
We now discuss the results focusing on three main aspects: (i) levels of expert-like thinking, (ii) differences between students' personal perspective and students' perspective of experts and (iii) changes upon instruction.

\subsubsection{\label{sec:discussionlevels}Levels of agreement with experts}
The results of the data analysis in Fig. \ref{fig1} and \ref{fig2} indicate that there is a clear difference between the level of 'expert-like' thinking about experimental physics of students at Potsdam and at other (mostly US) institutions. Such a difference is also observed when we look at students' opinions about expert physicists. 
We conclude that further efforts to improve the views of students on the nature of experimental physics at the University of Potsdam is a worthwhile endeavour. Moreover, these results lead to additional question, such as are these differences arising from a lack of understanding of the nature of experimental physics of German students in general as compared to other educational environments or is this result specific to the University of Potsdam. To answer this important question, we need to further investigate laboratory courses at other German-speaking institutions using the GE-CLASS. The GE-CLASS web-based system described in this paper allows this as a next step.
Regarding the data presented in Fig. \ref{fig2}, we conclude, from an instructional point of view, that we need to put particular effort in facilitating the aspects probed by the items in the left side of this figure. Those are the aspects that show the lowest agreement with the ER.

Looking at the ranking of the survey items in Fig. \ref{fig3}, we conclude that the aspects with the most/least 'expert-like' views of experimental physics do not depend on the educational environments (countries) investigated in this study.  These findings are informative for both instructors and the physics education community, as we work to broaden educational improvement efforts beyond our own countries.

\subsubsection{\label{sec:discussionstudentscontradictions}Students' contradictions}
As discussed in the previous section and in accordance to the literature \cite{Gray08}, we found in this study for physics majors (for both E-/GE-CLASS) that the overall agreement with experts for the YOU-questions is lower than the one for the EXPERT-questions. Students often know what are experts’ views and attitudes in the laboratory, but do not personally practice those perspectives when working in laboratory courses. This possible contradiction might arise from a perception of students that laboratory courses are not authentic experimental physics experiences or the type of activities they engage in are not well aligned with authentic practice. 

From Fig. \ref{fig2}, we find that the aspects for which we measured the largest differences between personal and expert views are the same for the two data sets GE/E-CLASS (see for example items 29, 6, 14, 10 and 5) . The reason why we observed the largest gaps for the same items is unknown, but may be due to a few reasons. For example, the discrepancy for question 5 (\textit{‘ Calculating uncertainties usually helps me understand my results better’}) can possibly be explained considering previous studies on students' understanding of  measurements uncertainty, which show a variety of student challenges with these concepts and practices in courses in many different countries \cite{Pillay08, Allie, Pollard2017, Pollard2020}.

For the items where students hold different views than their perception of experts, we must look at the practices and structures of our courses to posit why these may be different and how we might address this gap. 

\subsubsection{\label{sec:discussionchangesuponinstruction}Changes upon instruction}
We now discuss changes upon instruction for the E-/GE-CLASS data.
As outlined in the previous section, we find for both YOU- and EXPERTS-questions statistically significant shifts between pre- and post-surveys (Fig. \ref{fig1}). Our transformed curriculum has therefore a small, but positive, influence on students’ views and attitudes towards experimental physics. By comparing those results with the E-CLASS results, in which no significant shift is observed in Fig. \ref{fig1}, we argue that this small shift may be due to the fact that we identified items covered on the survey as an explicit goals of our lab class. 
For example, the largest positive shifts for questions Q29 and Q17 (Q29 \textit{‘If I don't have clear directions for analyzing data, I am not sure how to choose an appropriate analysis method’} and Q17 \textit{‘When I encounter difficulties in the lab, my first step is to ask an expert, like the instructor’}) were addressed in the course transformation.  We started the transformation by adapting the student laboratory manuals to remove the precise instructions on how to perform the experiment and data analysis and instead added self-guiding questions to help students reflect on how to solve problems or how to perform the data analysis. With this approach, we allowed for student decision making in the lab. This practice may have led to positive shifts on these two questions. Further qualitative research (classroom observations and student interviews) would need to be conducted to confirm this idea. We do note that the positive shift for Q17 may have been impacted by the pandemic and the resulting course modality. A previous study looked at how E-CLASS scores shifted from pre-pandemic to Fall 2020 and found a strong positive shift for Q17, which was attributed to, at least in part, the remote nature of the labs where access to instructor help may not have been as readily available \cite{FoxEclasssPandemic}.

On the other end, the largest negative shifts for the GE-CLASS are observed for YOU-question Q18 (\textit{‘Communicating scientific results to peers is a valuable part of doing physics experiments.’} and Q19 (\textit{‘Working in a group is an important part of doing physics experiments’}), which are related to communication and working in a group. We wonder if this effect is due to changes in the instructions we had to make due to the Covid19 pandemic ($N=42$ pre- and $N=83$ post-Covid19), for which group work was less encouraged compared to before the pandemic and students had to work alone a lot. Regardless, we are now working on specifically fostering communication and team-work by activities that emphasize the importance and advantages of team-work and peer discussions in an authentic manner.

\subsection{\label{sec:limiations} Limitations}

There are a few limitations to the work presented here. First, we note that a large fraction of the data of the GE-CLASS sample was obtained during the COVID-19 pandemic. This might have influenced our results in unpredictable ways, including impacts of the stress and uncertainty of the pandemic on the students. For various reasons, the response rates are low for the GE-CLASS. Thus, the small sample of students from each class may be biased, as students who were experiencing high levels of stress may have chosen not to complete the surveys.

Second, in the case of the E-CLASS, students are often given an incentive for doing the survey, such as a small amount of course credit for completion. Since this incentive was missing in the GE-CLASS, it is possible that students that were more engaged in the course and its contents answered the survey and thus represent a biased sample.  We believe that this form of possible bias does not invalidate the results, but must be taken into account when considering the stronger shifts upon instruction of the GE-CLASS.

Finally, this low response rate led to a relatively small dataset to analyse for the GE-CLASS.  In an ideal world, we would wait to collect enough data from the University of Potsdam to be able to split the data into FY and BFY datasets. However, due to the pandemic and low response rates caused by adhering to GDPR rules, this would take many years. Critically, this would delay the broad scale use of GE-CLASS across Germany and, thus, delay improvements in German lab classes along this one important dimension.

\section{\label{sec:conclusions}Conclusions}
We presented the process of how the E-CLASS was adapted to the German-speaking context. It is now available for instructors to assess laboratory classes in German-speaking institutions through an online automated system. This system will allow instructors and researchers to gain insights into the influence of laboratory classes on students' epistemology and views about experimental physics in German-speaking institutions, a field that has been unexplored until now.
    
First results using the GE-CLASS have been obtained at the University of Potsdam, as an example use of the GE-CLASS. The results give interesting insights on how the GE-CLASS can assess laboratory courses and guide the process of course transformations. This study using the GE-CLASS allowed us to make a first comparison with the results obtained from the E-CLASS. From this comparison, we found that there are similarities, but also differences between instructional contexts. For example, in both E-/GE-CLASS results, students have a good understanding about what experts think about experimental physics, but they do not always share those views during their own laboratory class. 
The aspects of experimental physics for which students have the most and least expert-like personal views are the same for the E-CLASS and GE-CLASS, indicating that the attitudes towards experimental physics show similar characteristics even considering these different educational environments.
On the other hand, we found that the level of 'expert-thinking' is systematically lower for students who took the GE-CLASS at the University of Potsdam as compared to students who took the E-CLASS in the US, and that the effects upon instruction are larger for the GE-CLASS. The analysis of the cumulative expert likelihood distribution shows that students at Potsdam reached lower scores both pre and post instruction than students did on E-CLASS for the YOU-questions.   
These results indicate the need to further address students' epistemology and views about experimental physics at the University of Potsdam and leads to questions about if these findings would be similar if studies were carried out at other German institutions.

In the future, we hope to use the GE-CLASS to answer additional research questions using an expanded dataset collected from both the University of Potsdam and many additional institutions across Germany. Possible questions include: (i) To what extent do students develop habits of mind, experimental strategies, enthusiasm, and confidence in doing experimental physics in German-speaking laboratory courses broadly? (ii) Are different teaching approaches in German-speaking laboratory courses impacting students’ attitudes and beliefs about experimental physics? (iii) Are there any differences between the US and German-speaking results when we compare the results using the data from a broad selection of lab courses across German-speaking countries?  
The answers to these research questions will be of great value for the further education of physics students in German-speaking, as well as other European, laboratory courses.

\begin{acknowledgments}
We would like to thank J. Beck, M. Gühr, and M. Robinson for helping with the translation of the E-CLASS. We also thank M. Path and R. Zumpe for helping students taking the GE-CLASS survey during the laboratory courses. This project was funded by the Stifterverband and the Baden-Württemberg Stiftung and by National Science Foundation (Grant No. PHY-1734006).

\end{acknowledgments}

\appendix
\section{\label{sec:appendix}German Translation of the E-CLASS}

Here, we report a list of all E-CLASS items and their translation into German.
In the parenthesis at the end of the E-CLASS items, we indicate the answer of the ER. 'A' stands for AGREE and 'D' stands for DISAGREE.

\begin{itemize}
\item[] Q1: When doing an experiment, I try to understand how the experimental setup works. (ER: A)
    
Bei der Durchführung eines Experimentes versuche ich zu verstehen wie der Versuchsaufbau funktioniert. 

\item[] Q2: If I wanted to, I think I could be good at doing research. (ER: A)

Ich denke ich könnte gut in der Durchführung von Forschung sein, falls ich wollte.

\item[] Q3: When doing a physics experiment, I don’t think much about sources of systematic error. (ER: D)

Während der Durchführung eines physikalischen Experiments denke ich nicht viel über Quellen systematischer Fehler nach.

\item[] Q4: If I am communicating results from an experiment, my main goal is to create a report with the correct sections and formatting. (ER: D)

Wenn ich über die Ergebnisse eines Experiments berichte, ist mein Hauptziel einen Bericht mit korrekten Abschnitten und Formatierung zu erstellen.

\item[] Q5: Calculating uncertainties usually helps me understand my results better. (ER: A)

Das Berechnen von Messunsicherheiten hilft mir üblicherweise meine Ergebnisse besser zu verstehen.

\item[] Q6: Scientific journal articles are helpful for answering my own questions and designing experiments. (ER: A)

Artikel aus wissenschaftlichen Fachzeitschriften helfen mir meine eigenen Fragen zu beantworten und beim Entwurf von Experimenten.

\item[] Q7: I don’t enjoy doing physics experiments. (ER: D)

Ich führe physikalische Experimente ungern durch.

\item[] Q8: When doing an experiment, I try to understand the relevant equations. (ER: A)

Während ich ein Experiment durchführe, versuche ich die relevanten Gleichungen zu verstehen.

\item[] Q9: When I approach a new piece of lab equipment, I feel confident I can learn how to use it well enough for my purposes. (ER: A)

Wenn ich auf ein neues Laborgerät stoße, bin ich mir sicher, dass ich lernen kann, für meine Zwecke gut genug damit umzugehen.

\item[] Q10: Whenever I use a new measurement tool, I try to understand its performance limitations. (ER: A)

Immer wenn ich ein neues Messgerät nutze, versuche ich die Grenzen seiner Leistungsfähigkeit zu verstehen.

\item[] Q11: Computers are helpful for plotting and analyzing data. (ER: A)

Computer sind hilfreich zur graphischen Darstellung und Analyse von Daten.

\item[] Q12: I don’t need to understand how the measurement tools and sensors work in order to carry out an experiment. (ER: D)

Ich muss nicht verstehen, wie Sensoren und Messinstrumente funktionieren, um ein Experiment durchzuführen.

\item[] Q13: If I try hard enough I can succeed at doing physics experiments. (ER: A)

Wenn ich mich genug anstrenge, kann ich erfolgreich physikalische Experimente durchführen.

\item[] Q14: When doing an experiment I usually think up my own questions to investigate.  (ER: A)

Wenn ich ein Experiment durchführe überlege ich mir normalerweise meine eigenen Fragen zu erforschen.

\item[] Q15: Designing and building things is an important part of doing physics experiments. (ER: A)

Das Entwerfen und Aufbauen von Gegenständen ist ein wichtiger Teil von physikalischen Experimenten.

\item[] Q16: The primary purpose of doing a physics experiment is to confirm previously known results. (ER: D)

Der Hauptzweck eines physikalischen Experimentes besteht in der Bestätigung bereits bekannter Resultate.

\item[] Q17: When I encounter difficulties in the lab, my first step is to ask an expert, like the instructor. (ER: D)

Wenn ich auf Schwierigkeiten im Labor stoße ist mein erster Schritt, einen Experten, wie z.B. den Betreuer, zu fragen.

\item[] Q18: Communicating scientific results to peers is a valuable part of doing physics experiments. (ER: A)

Die Vermittlung von wissenschaftlichen Resultaten an Kommilitonen ist ein wertvoller Teil des physikalischen Experimentierens.

\item[] Q19: Working in a group is an important part of doing physics experiments. (ER: A)

Gruppenarbeit ist ein wichtiger Teil der Durchführung physikalischer Experimente.

\item[] Q20: I enjoy building things and working with my hands. (ER: A)

Ich baue gerne Sachen und arbeite gerne mit meinen Händen.

\item[] Q21: I am usually able to complete an experiment without understanding the equations and physics ideas that describe the system I am investigating. (ER: D)

Meistens kann ich ein Experiment abschließen, ohne dass ich die Gleichungen und die physikalischen Konzepte verstehe, die das untersuchte System beschreiben.

\item[] Q22: If I am communicating results from an experiment, my main goal is to make conclusions based on my data using scientific reasoning. (ER: A)

Für mich ist bei der Präsentation meiner experimentellen Ergebnisse am wichtigsten, dass die Schlussfolgerungen aus meinen Daten unter Nutzung wissenschaftlicher Überlegungen hervorkommen.

\item[] Q23: When I am doing an experiment, I try to make predictions to see if my results are reasonable. (ER: A)

Wenn ich ein Experiment durchführe, versuche ich Vorhersagen zu treffen, um zu sehen ob meine Resultate sinnvoll sind.

\item[] Q24: Nearly all students are capable of doing a physics experiment if they work at it. (ER: A)

Fast alle Studierenden sind fähig ein physikalisches Experiment durchzuführen wenn sie sich Mühe geben.

\item[] Q25: A common approach for fixing a problem with an experiment is to randomly change things until the problem goes away. (ER: D)

Um ein Problem am Experiment zu beheben ändert man üblicherweise zufällig Sachen bis das Problem verschwindet.

\item[] Q26: It is helpful to understand the assumptions that go into
making predictions. (ER: A)

Es ist hilfreich die vorausgesetzten Annahmen zu verstehen, die in Vorhersagen einfließen.

\item[] Q27: When doing an experiment, I just follow the instructions without thinking about their purpose. (ER: D)

Während ich ein Experiment durchführe folge ich einfach der Anweisung ohne über ihren Zweck nachzudenken.

\item[] Q28: I do not expect doing an experiment to help my understanding of physics. (ER: D)

Ich erwarte nicht, dass die Durchführung eines Experiments mir beim Verstehen von Physik hilft.

\item[] Q29: If I don’t have clear directions for analyzing data, I am not sure how to choose an appropriate analysis method. (ER: D)

Ohne klare Anweisungen zur Datenanalyse bin ich mir unsicher über die angemessene Analysemethode.

\item[] Q30: Physics experiments contribute to the growth of scientific knowledge. (ER: A)

Physikalische Experimente tragen zum Wachstum des wissenschaftlichen Wissensschatzes bei.

\end{itemize}

The following questions are asked only in the post test:

\begin{itemize}

\item[] How important for earning a good grade in
   this class was...
   
   Wie wichtig für eine gute Note im Praktikum war…:
   
\item[] ... understanding how the experimental setup works?

... es zu verstehen, wie der experimentelle Aufbau funktioniert?

\item[] ... thinking about sources of systematic error?

... das Nachdenken über systematische Fehlerquellen?

\item[] ... communicating results with the correct sections and formatting?

... Ergebnisse mit korrekter Gliederung und Formatierung zu vermitteln?

\item[] ... calculating uncertainties to better understand my results?

... die Berechnung von Messunsicherheiten zum besseren Verständnis meiner Ergebnisse?

\item[] ... reading scientific journal articles?

... das Lesen von Artikeln in wissenschaftlichen Fachzeitschriften?

\item[] ... understanding the relevant equations?

... das Verständnis der relevanten Gleichungen?

\item[] ... learning to use a new piece of laboratory 
equipment?

... das Erlernen der Nutzung von neuen Laborinstrumenten?

\item[] ... understanding the performance limitations of the  measurement tools?

... das Verstehen der Leistungsgrenzen der Meßinstrumente?

\item[] ... using a computer for plotting and analyzing data?

... die Nutzung eines Computers zur graphischen Darstellung und Analyse der Daten?

\item[] ... understanding how the measurement tools and 
sensors work?

... es zu verstehen wie die Messinstrumente und Sensoren funktionieren?

\item[] ... thinking up my own questions to investigate?

... meine eigenen Fragen zur Erforschung zu überlegen?

\item[] ... designing and building things?

... der Entwurf und Aufbau von Gegenständen?

\item[] ... confirming previously known results?

... die Bestätigung bereits bekannter Resultate?

\item[] ... overcoming difficulties without the instructor’s  help?

... die Überwindung von Schwierigkeiten ohne die Hilfe des Betreuers?

\item[] ... communicating scientific results to peers?

... die Vermittlung wissenschaftlicher Ergebnisse an Kommilitonen?

\item[] ... working in a group?

... die Arbeit in der Gruppe?

\item[] ... understanding the equations and physics ideas that
describe the system I am investigating?

... das Verständnis der Gleichungen und der physikalischen Konzepte die das untersuchte System beschreiben?

\item[] ... making conclusions based on data using scientific reasoning?

... basierend auf Daten und wissenschaftlicher Denkweise, Rückschlüsse zu ziehen?

\item[] ... making predictions to see if my results are reasonable?

... es, Vorhersage zu treffen um zu sehen ob meine Ergebnisse sinnvoll sind?

\item[] ... randomly changing things to fix a problem with the experiment?

... zufällig Sachen am Experiment zu ändern bis ein Problem am Experiment verschwindet?

\item[] ... understanding the approximations and simplifications that are included in theoretical predictions?

... das Verständnis der Näherungen und Vereinfachungen die in die theoretischen Vorhersagen eingehen?

\item[] ... thinking about the purpose of the instructions in the lab guide?

... das Nachdenken über den Zweck der Anleitungen in der Versuchsanleitung?

\item[] ... choosing an appropriate method for analyzing data (without explicit direction)?

... die Auswahl einer angemessenen Methode zur Datenanalyse (ohne explizite Anleitung)?

\end{itemize}

\providecommand{\noopsort}[1]{}\providecommand{\singleletter}[1]{#1}%


\begin{thebibliography}{56}%
\makeatletter
\providecommand \@ifxundefined [1]{%
 \@ifx{#1\undefined}
}%
\providecommand \@ifnum [1]{%
 \ifnum #1\expandafter \@firstoftwo
 \else \expandafter \@secondoftwo
 \fi
}%
\providecommand \@ifx [1]{%
 \ifx #1\expandafter \@firstoftwo
 \else \expandafter \@secondoftwo
 \fi
}%
\providecommand \natexlab [1]{#1}%
\providecommand \enquote  [1]{``#1''}%
\providecommand \bibnamefont  [1]{#1}%
\providecommand \bibfnamefont [1]{#1}%
\providecommand \citenamefont [1]{#1}%
\providecommand \href@noop [0]{\@secondoftwo}%
\providecommand \href [0]{\begingroup \@sanitize@url \@href}%
\providecommand \@href[1]{\@@startlink{#1}\@@href}%
\providecommand \@@href[1]{\endgroup#1\@@endlink}%
\providecommand \@sanitize@url [0]{\catcode `\\12\catcode `\$12\catcode
  `\&12\catcode `\#12\catcode `\^12\catcode `\_12\catcode `\%12\relax}%
\providecommand \@@startlink[1]{}%
\providecommand \@@endlink[0]{}%
\providecommand \url  [0]{\begingroup\@sanitize@url \@url }%
\providecommand \@url [1]{\endgroup\@href {#1}{\urlprefix }}%
\providecommand \urlprefix  [0]{URL }%
\providecommand \Eprint [0]{\href }%
\providecommand \doibase [0]{https://doi.org/}%
\providecommand \selectlanguage [0]{\@gobble}%
\providecommand \bibinfo  [0]{\@secondoftwo}%
\providecommand \bibfield  [0]{\@secondoftwo}%
\providecommand \translation [1]{[#1]}%
\providecommand \BibitemOpen [0]{}%
\providecommand \bibitemStop [0]{}%
\providecommand \bibitemNoStop [0]{.\EOS\space}%
\providecommand \EOS [0]{\spacefactor3000\relax}%
\providecommand \BibitemShut  [1]{\csname bibitem#1\endcsname}%
\let\auto@bib@innerbib\@empty
\bibitem [{\citenamefont {Trumper}(2003)}]{Trumper03}%
  \BibitemOpen
  \bibfield  {author} {\bibinfo {author} {\bibfnamefont {R.}~\bibnamefont
  {Trumper}},\ }\bibfield  {title} {\bibinfo {title} {The physics
  laboratory—a historical overview and future perspectives},\ }\href@noop {}
  {\bibfield  {journal} {\bibinfo  {journal} {Sci. Educ.}\ }\textbf {\bibinfo
  {volume} {12}},\ \bibinfo {pages} {645} (\bibinfo {year} {2003})}\BibitemShut
  {NoStop}%
\bibitem [{AAP(2015)}]{AAPT15}%
  \BibitemOpen
  \bibfield  {title} {\bibinfo {title} {American association physics teachers
  committee on laboratories},\ }\href@noop {} {\bibfield  {journal} {\bibinfo
  {journal} {AAPT Recommendations for the Undergraduate Physics Laboratory
  Curriculum}\ } (\bibinfo {year} {2015})}\BibitemShut {NoStop}%
\bibitem [{\citenamefont {AAPT}(1998)}]{AAPT98}%
  \BibitemOpen
  \bibfield  {author} {\bibinfo {author} {\bibnamefont {AAPT}},\ }\bibfield
  {title} {\bibinfo {title} {Goals of the introductory physics laboratory},\
  }\href@noop {} {\bibfield  {journal} {\bibinfo  {journal} {Am. J. Phys.}\
  }\textbf {\bibinfo {volume} {66}},\ \bibinfo {pages} {483} (\bibinfo {year}
  {1998})}\BibitemShut {NoStop}%
\bibitem [{\citenamefont {Nagel}\ \emph {et~al.}(2018)\citenamefont {Nagel},
  \citenamefont {Scholz},\ and\ \citenamefont {Weber}}]{Nagel18}%
  \BibitemOpen
  \bibfield  {author} {\bibinfo {author} {\bibfnamefont {C.}~\bibnamefont
  {Nagel}}, \bibinfo {author} {\bibfnamefont {R.}~\bibnamefont {Scholz}},\ and\
  \bibinfo {author} {\bibfnamefont {K.}~\bibnamefont {Weber}},\ }\bibfield
  {title} {\bibinfo {title} {Umfrage zu den {L}ehr/{L}ernzielen in
  physikalischen {P}raktika},\ }\href@noop {} {\bibfield  {journal} {\bibinfo
  {journal} {PhyDid B - DPG Frühjahrstagung}\ } (\bibinfo {year}
  {2018})}\BibitemShut {NoStop}%
\bibitem [{\citenamefont {Holmes}\ and\ \citenamefont
  {Lewandowski}(2020)}]{landscape}%
  \BibitemOpen
  \bibfield  {author} {\bibinfo {author} {\bibfnamefont {N.~G.}\ \bibnamefont
  {Holmes}}\ and\ \bibinfo {author} {\bibfnamefont {H.~J.}\ \bibnamefont
  {Lewandowski}},\ }\bibfield  {title} {\bibinfo {title} {Investigating the
  landscape of physics laboratory instruction across north america},\ }\href
  {https://doi.org/10.1103/PhysRevPhysEducRes.16.020162} {\bibfield  {journal}
  {\bibinfo  {journal} {Phys. Rev. Phys. Educ. Res.}\ }\textbf {\bibinfo
  {volume} {16}},\ \bibinfo {pages} {020162} (\bibinfo {year}
  {2020})}\BibitemShut {NoStop}%
\bibitem [{\citenamefont {Wieman}\ and\ \citenamefont
  {Holmes}(2015)}]{Wieman15}%
  \BibitemOpen
  \bibfield  {author} {\bibinfo {author} {\bibfnamefont {C.}~\bibnamefont
  {Wieman}}\ and\ \bibinfo {author} {\bibfnamefont {N.~G.}\ \bibnamefont
  {Holmes}},\ }\bibfield  {title} {\bibinfo {title} {Measuring the impact of an
  instructional laboratory on the learning of introductory physics},\
  }\href@noop {} {\bibfield  {journal} {\bibinfo  {journal} {Am. J. Phys.}\
  }\textbf {\bibinfo {volume} {83}},\ \bibinfo {pages} {972} (\bibinfo {year}
  {2015})}\BibitemShut {NoStop}%
\bibitem [{\citenamefont {Holmes}\ and\ \citenamefont
  {Wieman}(2018)}]{Wieman18}%
  \BibitemOpen
  \bibfield  {author} {\bibinfo {author} {\bibfnamefont {N.~G.}\ \bibnamefont
  {Holmes}}\ and\ \bibinfo {author} {\bibfnamefont {C.}~\bibnamefont
  {Wieman}},\ }\bibfield  {title} {\bibinfo {title} {Introductory physics labs:
  We can do better},\ }\href@noop {} {\bibfield  {journal} {\bibinfo  {journal}
  {Physics Today}\ }\textbf {\bibinfo {volume} {71}},\ \bibinfo {pages} {645}
  (\bibinfo {year} {2018})}\BibitemShut {NoStop}%
\bibitem [{\citenamefont {Etkina}\ \emph {et~al.}(2010)\citenamefont {Etkina},
  \citenamefont {Karelina}, \citenamefont {Ruibal-Villasenor}, \citenamefont
  {Rosengrant}, \citenamefont {Jordan},\ and\ \citenamefont
  {Hmelo-Silver}}]{Etk10}%
  \BibitemOpen
  \bibfield  {author} {\bibinfo {author} {\bibfnamefont {E.}~\bibnamefont
  {Etkina}}, \bibinfo {author} {\bibfnamefont {A.}~\bibnamefont {Karelina}},
  \bibinfo {author} {\bibfnamefont {M.}~\bibnamefont {Ruibal-Villasenor}},
  \bibinfo {author} {\bibfnamefont {D.}~\bibnamefont {Rosengrant}}, \bibinfo
  {author} {\bibfnamefont {R.}~\bibnamefont {Jordan}},\ and\ \bibinfo {author}
  {\bibfnamefont {C.}~\bibnamefont {Hmelo-Silver}},\ }\bibfield  {title}
  {\bibinfo {title} {Design and reflection help students develop scientific
  abilities: Learning in introductory physics laboratories},\ }\href@noop {}
  {\bibfield  {journal} {\bibinfo  {journal} {Journal of the Learning
  Sciences}\ }\textbf {\bibinfo {volume} {19}},\ \bibinfo {pages} {54}
  (\bibinfo {year} {2010})}\BibitemShut {NoStop}%
\bibitem [{\citenamefont {Pillay}\ \emph {et~al.}(2008)\citenamefont {Pillay},
  \citenamefont {Buffler}, \citenamefont {Lubben},\ and\ \citenamefont
  {Allie}}]{Pillay08}%
  \BibitemOpen
  \bibfield  {author} {\bibinfo {author} {\bibfnamefont {S.}~\bibnamefont
  {Pillay}}, \bibinfo {author} {\bibfnamefont {A.}~\bibnamefont {Buffler}},
  \bibinfo {author} {\bibfnamefont {F.}~\bibnamefont {Lubben}},\ and\ \bibinfo
  {author} {\bibfnamefont {S.}~\bibnamefont {Allie}},\ }\bibfield  {title}
  {\bibinfo {title} {Effectiveness of a {GUM}-compliant course for teaching
  measurement in the introductory physics laboratory},\ }\href@noop {}
  {\bibfield  {journal} {\bibinfo  {journal} {Eur. J. Phys.}\ }\textbf
  {\bibinfo {volume} {29}},\ \bibinfo {pages} {647} (\bibinfo {year}
  {2008})}\BibitemShut {NoStop}%
\bibitem [{\citenamefont {Volkwyn}\ \emph {et~al.}(2008)\citenamefont
  {Volkwyn}, \citenamefont {Allie}, \citenamefont {Buffler},\ and\
  \citenamefont {Lubben}}]{Volkwyn08}%
  \BibitemOpen
  \bibfield  {author} {\bibinfo {author} {\bibfnamefont {T.~S.}\ \bibnamefont
  {Volkwyn}}, \bibinfo {author} {\bibfnamefont {S.}~\bibnamefont {Allie}},
  \bibinfo {author} {\bibfnamefont {A.}~\bibnamefont {Buffler}},\ and\ \bibinfo
  {author} {\bibfnamefont {F.}~\bibnamefont {Lubben}},\ }\bibfield  {title}
  {\bibinfo {title} {Impact of a conventional introductory laboratory course on
  the understanding of measurement},\ }\href@noop {} {\bibfield  {journal}
  {\bibinfo  {journal} {Phys. Rev. ST Phys. Educ. Res.}\ }\textbf {\bibinfo
  {volume} {4}},\ \bibinfo {pages} {010108} (\bibinfo {year}
  {2008})}\BibitemShut {NoStop}%
\bibitem [{\citenamefont {Zwickl}\ \emph {et~al.}(2013)\citenamefont {Zwickl},
  \citenamefont {Finkelstein},\ and\ \citenamefont {Lewandowski}}]{Zwickl13}%
  \BibitemOpen
  \bibfield  {author} {\bibinfo {author} {\bibfnamefont {B.~M.}\ \bibnamefont
  {Zwickl}}, \bibinfo {author} {\bibfnamefont {N.}~\bibnamefont
  {Finkelstein}},\ and\ \bibinfo {author} {\bibfnamefont {H.~J.}\ \bibnamefont
  {Lewandowski}},\ }\bibfield  {title} {\bibinfo {title} {The process of
  transforming an advanced lab course: Goals, curriculum, and assessments},\
  }\href@noop {} {\bibfield  {journal} {\bibinfo  {journal} {Am. J. Phys.}\
  }\textbf {\bibinfo {volume} {81}},\ \bibinfo {pages} {63} (\bibinfo {year}
  {2013})}\BibitemShut {NoStop}%
\bibitem [{\citenamefont {Zwickl}\ \emph {et~al.}(2014)\citenamefont {Zwickl},
  \citenamefont {Hirokawa}, \citenamefont {Finkelstein},\ and\ \citenamefont
  {Lewandowski}}]{Zwickl14}%
  \BibitemOpen
  \bibfield  {author} {\bibinfo {author} {\bibfnamefont {B.~M.}\ \bibnamefont
  {Zwickl}}, \bibinfo {author} {\bibfnamefont {T.}~\bibnamefont {Hirokawa}},
  \bibinfo {author} {\bibfnamefont {N.}~\bibnamefont {Finkelstein}},\ and\
  \bibinfo {author} {\bibfnamefont {H.~J.}\ \bibnamefont {Lewandowski}},\
  }\bibfield  {title} {\bibinfo {title} {Epistemology and expectations survey
  about experimental physics: Development and initial results},\ }\href@noop {}
  {\bibfield  {journal} {\bibinfo  {journal} {Phys. Rev. ST Phys. Educ. Res.}\
  }\textbf {\bibinfo {volume} {10}},\ \bibinfo {pages} {010120} (\bibinfo
  {year} {2014})}\BibitemShut {NoStop}%
\bibitem [{\citenamefont {Wilcox}\ and\ \citenamefont
  {Lewandowski}(2018)}]{WilcoxLew18}%
  \BibitemOpen
  \bibfield  {author} {\bibinfo {author} {\bibfnamefont {B.~R.}\ \bibnamefont
  {Wilcox}}\ and\ \bibinfo {author} {\bibfnamefont {H.~J.}\ \bibnamefont
  {Lewandowski}},\ }\bibfield  {title} {\bibinfo {title} {A summary of
  research-based assessment of students' beliefs about the nature of
  experimental physics},\ }\href@noop {} {\bibfield  {journal} {\bibinfo
  {journal} {American Journal of Physics}\ }\textbf {\bibinfo {volume} {86}},\
  \bibinfo {pages} {212} (\bibinfo {year} {2018})}\BibitemShut {NoStop}%
\bibitem [{\citenamefont {Adams}\ \emph {et~al.}(2006)\citenamefont {Adams},
  \citenamefont {Perkins}, \citenamefont {Podolefsky}, \citenamefont {Dubson},
  \citenamefont {Finkelstein},\ and\ \citenamefont {Wieman}}]{Adams06}%
  \BibitemOpen
  \bibfield  {author} {\bibinfo {author} {\bibfnamefont {W.~K.}\ \bibnamefont
  {Adams}}, \bibinfo {author} {\bibfnamefont {K.~K.}\ \bibnamefont {Perkins}},
  \bibinfo {author} {\bibfnamefont {N.~S.}\ \bibnamefont {Podolefsky}},
  \bibinfo {author} {\bibfnamefont {M.}~\bibnamefont {Dubson}}, \bibinfo
  {author} {\bibfnamefont {N.~D.}\ \bibnamefont {Finkelstein}},\ and\ \bibinfo
  {author} {\bibfnamefont {C.~E.}\ \bibnamefont {Wieman}},\ }\bibfield  {title}
  {\bibinfo {title} {New instrument for measuring student beliefs about physics
  and learning physics: The {C}olorado {L}earning {A}ttitudes about {S}cience
  {S}urvey},\ }\href@noop {} {\bibfield  {journal} {\bibinfo  {journal} {Phys.
  Rev. ST Phys. Educ. Res.}\ }\textbf {\bibinfo {volume} {2}},\ \bibinfo
  {pages} {010101} (\bibinfo {year} {2006})}\BibitemShut {NoStop}%
\bibitem [{\citenamefont {Redish}\ \emph {et~al.}(1998)\citenamefont {Redish},
  \citenamefont {Saul}, ,\ and\ \citenamefont {Steinberg}}]{Redish98}%
  \BibitemOpen
  \bibfield  {author} {\bibinfo {author} {\bibfnamefont {E.~F.}\ \bibnamefont
  {Redish}}, \bibinfo {author} {\bibfnamefont {J.~M.}\ \bibnamefont {Saul}}, ,\
  and\ \bibinfo {author} {\bibfnamefont {R.~N.}\ \bibnamefont {Steinberg}},\
  }\bibfield  {title} {\bibinfo {title} {Student expectations in introductory
  physics},\ }\href@noop {} {\bibfield  {journal} {\bibinfo  {journal} {Am. J.
  Phys.}\ }\textbf {\bibinfo {volume} {66}},\ \bibinfo {pages} {212} (\bibinfo
  {year} {1998})}\BibitemShut {NoStop}%
\bibitem [{\citenamefont {Perkins}\ \emph {et~al.}(2005)\citenamefont
  {Perkins}, \citenamefont {Adams}, \citenamefont {Pollock}, \citenamefont
  {Finkelstein},\ and\ \citenamefont {Wieman}}]{Perkins05}%
  \BibitemOpen
  \bibfield  {author} {\bibinfo {author} {\bibfnamefont {K.}~\bibnamefont
  {Perkins}}, \bibinfo {author} {\bibfnamefont {W.}~\bibnamefont {Adams}},
  \bibinfo {author} {\bibfnamefont {S.}~\bibnamefont {Pollock}}, \bibinfo
  {author} {\bibfnamefont {N.}~\bibnamefont {Finkelstein}},\ and\ \bibinfo
  {author} {\bibfnamefont {C.}~\bibnamefont {Wieman}},\ }\bibfield  {title}
  {\bibinfo {title} {Correlating student beliefs with student learning using
  the {C}olorado {L}earning {A}ttitudes about {S}cience {S}urvey},\ }\href@noop
  {} {\bibfield  {journal} {\bibinfo  {journal} {AIP Conf. Proc.}\ }\textbf
  {\bibinfo {volume} {790}},\ \bibinfo {pages} {61} (\bibinfo {year}
  {2005})}\BibitemShut {NoStop}%
\bibitem [{\citenamefont {Lising}\ and\ \citenamefont {Elby}(2005)}]{Lising05}%
  \BibitemOpen
  \bibfield  {author} {\bibinfo {author} {\bibfnamefont {L.}~\bibnamefont
  {Lising}}\ and\ \bibinfo {author} {\bibfnamefont {A.}~\bibnamefont {Elby}},\
  }\bibfield  {title} {\bibinfo {title} {The impact of epistemology on
  learning: A case study from introductory physics},\ }\href@noop {} {\bibfield
   {journal} {\bibinfo  {journal} {Am. J. Phys.}\ }\textbf {\bibinfo {volume}
  {73}},\ \bibinfo {pages} {372} (\bibinfo {year} {2005})}\BibitemShut
  {NoStop}%
\bibitem [{\citenamefont {Handelsman}\ \emph {et~al.}(2004)\citenamefont
  {Handelsman}, \citenamefont {Ebert-May}, \citenamefont {Beichner},
  \citenamefont {Bruns}, \citenamefont {Chang}, \citenamefont {DeHaan},
  \citenamefont {Gentile}, \citenamefont {Lauffer}, \citenamefont {Stewart},
  \citenamefont {Tilghman},\ and\ \citenamefont {Wood}}]{Handelsman04}%
  \BibitemOpen
  \bibfield  {author} {\bibinfo {author} {\bibfnamefont {J.}~\bibnamefont
  {Handelsman}}, \bibinfo {author} {\bibfnamefont {D.}~\bibnamefont
  {Ebert-May}}, \bibinfo {author} {\bibfnamefont {R.}~\bibnamefont {Beichner}},
  \bibinfo {author} {\bibfnamefont {P.}~\bibnamefont {Bruns}}, \bibinfo
  {author} {\bibfnamefont {A.}~\bibnamefont {Chang}}, \bibinfo {author}
  {\bibfnamefont {R.}~\bibnamefont {DeHaan}}, \bibinfo {author} {\bibfnamefont
  {J.}~\bibnamefont {Gentile}}, \bibinfo {author} {\bibfnamefont
  {S.}~\bibnamefont {Lauffer}}, \bibinfo {author} {\bibfnamefont
  {J.}~\bibnamefont {Stewart}}, \bibinfo {author} {\bibfnamefont {S.~M.}\
  \bibnamefont {Tilghman}},\ and\ \bibinfo {author} {\bibfnamefont {W.~B.}\
  \bibnamefont {Wood}},\ }\bibfield  {title} {\bibinfo {title} {Scientific
  {T}eaching},\ }\href@noop {} {\bibfield  {journal} {\bibinfo  {journal}
  {Science}\ }\textbf {\bibinfo {volume} {304}},\ \bibinfo {pages} {5670}
  (\bibinfo {year} {2004})}\BibitemShut {NoStop}%
\bibitem [{\citenamefont {Etkina}\ and\ \citenamefont {Heuvelen}()}]{Etk01}%
  \BibitemOpen
  \bibfield  {author} {\bibinfo {author} {\bibfnamefont {E.}~\bibnamefont
  {Etkina}}\ and\ \bibinfo {author} {\bibfnamefont {A.~V.}\ \bibnamefont
  {Heuvelen}},\ }\bibfield  {title} {\bibinfo {title} {Investigative science
  learning environment: Using the processes of science and cognitive strategies
  to learn physics},\ }\href@noop {} {\bibinfo  {journal} {in 2001 PERC
  Proceedings, Rochester, NY, July 25-26, 2001, edited by K. Cummings, S.
  Franklin, and J. Marx (AAPT, College Park, MD, 2001)}\ }\BibitemShut
  {NoStop}%
\bibitem [{\citenamefont {Etkina}\ \emph {et~al.}(2006)\citenamefont {Etkina},
  \citenamefont {Heuvelen}, \citenamefont {White-Brahmia}, \citenamefont
  {Brookes}, \citenamefont {Gentile}, \citenamefont {Murthy}, \citenamefont
  {Rosengrant},\ and\ \citenamefont {Warren}}]{Etk06}%
  \BibitemOpen
\bibfield  {journal} {  }\bibfield  {author} {\bibinfo {author} {\bibfnamefont
  {E.}~\bibnamefont {Etkina}}, \bibinfo {author} {\bibfnamefont {A.~V.}\
  \bibnamefont {Heuvelen}}, \bibinfo {author} {\bibfnamefont {S.}~\bibnamefont
  {White-Brahmia}}, \bibinfo {author} {\bibfnamefont {D.~T.}\ \bibnamefont
  {Brookes}}, \bibinfo {author} {\bibfnamefont {M.}~\bibnamefont {Gentile}},
  \bibinfo {author} {\bibfnamefont {S.}~\bibnamefont {Murthy}}, \bibinfo
  {author} {\bibfnamefont {D.}~\bibnamefont {Rosengrant}},\ and\ \bibinfo
  {author} {\bibfnamefont {A.}~\bibnamefont {Warren}},\ }\bibfield  {title}
  {\bibinfo {title} {Scientific abilities and their assessment},\ }\href@noop
  {} {\bibfield  {journal} {\bibinfo  {journal} {Phys. Rev. ST Phys. Educ.
  Res.}\ }\textbf {\bibinfo {volume} {2}},\ \bibinfo {pages} {020103} (\bibinfo
  {year} {2006})}\BibitemShut {NoStop}%
\bibitem [{\citenamefont {Karelina}\ and\ \citenamefont
  {Etkina}(2007)}]{Etk07}%
  \BibitemOpen
  \bibfield  {author} {\bibinfo {author} {\bibfnamefont {A.}~\bibnamefont
  {Karelina}}\ and\ \bibinfo {author} {\bibfnamefont {E.}~\bibnamefont
  {Etkina}},\ }\bibfield  {title} {\bibinfo {title} {Acting like a physicist:
  Student {A}pproach {S}tudy to {E}xperimental {D}esign},\ }\href@noop {}
  {\bibfield  {journal} {\bibinfo  {journal} {Phys. Rev. ST Phys. Educ. Res.}\
  }\textbf {\bibinfo {volume} {3}},\ \bibinfo {pages} {020106} (\bibinfo {year}
  {2007})}\BibitemShut {NoStop}%
\bibitem [{\citenamefont {Holmes}\ \emph {et~al.}(2015)\citenamefont {Holmes},
  \citenamefont {Wieman},\ and\ \citenamefont {Bonn}}]{Holmes15}%
  \BibitemOpen
  \bibfield  {author} {\bibinfo {author} {\bibfnamefont {N.~G.}\ \bibnamefont
  {Holmes}}, \bibinfo {author} {\bibfnamefont {C.~E.}\ \bibnamefont {Wieman}},\
  and\ \bibinfo {author} {\bibfnamefont {D.~A.}\ \bibnamefont {Bonn}},\
  }\bibfield  {title} {\bibinfo {title} {Teaching critical thinking},\
  }\href@noop {} {\bibfield  {journal} {\bibinfo  {journal} {Proc. Natl. Acad.
  Sci. U.S.A.}\ }\textbf {\bibinfo {volume} {112}},\ \bibinfo {pages} {11199}
  (\bibinfo {year} {2015})}\BibitemShut {NoStop}%
\bibitem [{\citenamefont {Zwickl}\ \emph {et~al.}(2015)\citenamefont {Zwickl},
  \citenamefont {D.~Hu},\ and\ \citenamefont {Lewandowski}}]{Zwickl15}%
  \BibitemOpen
  \bibfield  {author} {\bibinfo {author} {\bibfnamefont {B.~M.}\ \bibnamefont
  {Zwickl}}, \bibinfo {author} {\bibfnamefont {N.~F.}\ \bibnamefont {D.~Hu}},\
  and\ \bibinfo {author} {\bibfnamefont {H.~J.}\ \bibnamefont {Lewandowski}},\
  }\bibfield  {title} {\bibinfo {title} {Model-based reasoning in the physics
  laboratory: Framework and initial results},\ }\href@noop {} {\bibfield
  {journal} {\bibinfo  {journal} {Phys. Rev. ST Phys. Educ. Res.}\ }\textbf
  {\bibinfo {volume} {11}},\ \bibinfo {pages} {020113} (\bibinfo {year}
  {2015})}\BibitemShut {NoStop}%
\bibitem [{\citenamefont {Dounas-Frazer}\ \emph {et~al.}(2020)\citenamefont
  {Dounas-Frazer}, \citenamefont {Johnson}, \citenamefont {Park}, \citenamefont
  {Stanley},\ and\ \citenamefont {Lewandowski}}]{Frazer2020}%
  \BibitemOpen
  \bibfield  {author} {\bibinfo {author} {\bibfnamefont {D.}~\bibnamefont
  {Dounas-Frazer}}, \bibinfo {author} {\bibfnamefont {K.~S.}\ \bibnamefont
  {Johnson}}, \bibinfo {author} {\bibfnamefont {S.~E.}\ \bibnamefont {Park}},
  \bibinfo {author} {\bibfnamefont {J.~T.}\ \bibnamefont {Stanley}},\ and\
  \bibinfo {author} {\bibfnamefont {H.~J.}\ \bibnamefont {Lewandowski}},\
  }\bibfield  {title} {\bibinfo {title} {Student perceptions of laboratory
  classroom activities and experimental physics practice},\ }in\ \href@noop {}
  {\emph {\bibinfo {booktitle} {Physics Education Research Conference 2020}}},\
  \bibinfo {series and number} {PER Conference}\ (\bibinfo {address} {Virtual
  Conference},\ \bibinfo {year} {2020})\ pp.\ \bibinfo {pages}
  {131--136}\BibitemShut {NoStop}%
\bibitem [{phy()}]{physport}%
  \BibitemOpen
  \href@noop {} {\bibinfo {title}
  {\url{https://www.physport.org/assessments}}}\BibitemShut {NoStop}%
\bibitem [{\citenamefont {Day}\ and\ \citenamefont {Bonn}(2011)}]{Day11}%
  \BibitemOpen
  \bibfield  {author} {\bibinfo {author} {\bibfnamefont {J.}~\bibnamefont
  {Day}}\ and\ \bibinfo {author} {\bibfnamefont {D.}~\bibnamefont {Bonn}},\
  }\bibfield  {title} {\bibinfo {title} {Development of the concise data
  processing assessment},\ }\href@noop {} {\bibfield  {journal} {\bibinfo
  {journal} {Phys. Rev. ST Phys. Educ. Res.}\ }\textbf {\bibinfo {volume}
  {7}},\ \bibinfo {pages} {010114} (\bibinfo {year} {2011})}\BibitemShut
  {NoStop}%
\bibitem [{\citenamefont {Walsh}\ \emph {et~al.}(2019)\citenamefont {Walsh},
  \citenamefont {Quinn}, \citenamefont {Wieman},\ and\ \citenamefont
  {Holmes}}]{PLIC19}%
  \BibitemOpen
  \bibfield  {author} {\bibinfo {author} {\bibfnamefont {C.}~\bibnamefont
  {Walsh}}, \bibinfo {author} {\bibfnamefont {K.~N.}\ \bibnamefont {Quinn}},
  \bibinfo {author} {\bibfnamefont {C.}~\bibnamefont {Wieman}},\ and\ \bibinfo
  {author} {\bibfnamefont {N.~G.}\ \bibnamefont {Holmes}},\ }\bibfield  {title}
  {\bibinfo {title} {Quantifying critical thinking: Development and validation
  of the physics lab inventory of critical thinking},\ }\href
  {https://doi.org/10.1103/PhysRevPhysEducRes.15.010135} {\bibfield  {journal}
  {\bibinfo  {journal} {Phys. Rev. Phys. Educ. Res.}\ }\textbf {\bibinfo
  {volume} {15}},\ \bibinfo {pages} {010135} (\bibinfo {year}
  {2019})}\BibitemShut {NoStop}%
\bibitem [{\citenamefont {Theyßen}\ \emph {et~al.}(2013)\citenamefont
  {Theyßen}, \citenamefont {Dickmann}, \citenamefont {Neumann}, \citenamefont
  {Schecker},\ and\ \citenamefont {Eickhorst}}]{Theyssen16}%
  \BibitemOpen
  \bibfield  {author} {\bibinfo {author} {\bibfnamefont {H.}~\bibnamefont
  {Theyßen}}, \bibinfo {author} {\bibfnamefont {M.}~\bibnamefont {Dickmann}},
  \bibinfo {author} {\bibfnamefont {K.}~\bibnamefont {Neumann}}, \bibinfo
  {author} {\bibfnamefont {H.}~\bibnamefont {Schecker}},\ and\ \bibinfo
  {author} {\bibfnamefont {B.}~\bibnamefont {Eickhorst}},\ }\bibfield  {title}
  {\bibinfo {title} {Measuring experimental skills in large scale assessments:
  a simulation-based test instrument},\ }\href@noop {} {\bibfield  {journal}
  {\bibinfo  {journal} {Proceedings of the bi-annual conference of the European
  Science Education Research Conference (ESERA)}\ } (\bibinfo {year}
  {2013})}\BibitemShut {NoStop}%
\bibitem [{\citenamefont {Theyßen}\ \emph {et~al.}(2016)\citenamefont
  {Theyßen}, \citenamefont {Schecker}, \citenamefont {Neumann}, \citenamefont
  {Eickhorst},\ and\ \citenamefont {Dickmann}}]{Theyssen18}%
  \BibitemOpen
  \bibfield  {author} {\bibinfo {author} {\bibfnamefont {H.}~\bibnamefont
  {Theyßen}}, \bibinfo {author} {\bibfnamefont {H.}~\bibnamefont {Schecker}},
  \bibinfo {author} {\bibfnamefont {K.}~\bibnamefont {Neumann}}, \bibinfo
  {author} {\bibfnamefont {B.}~\bibnamefont {Eickhorst}},\ and\ \bibinfo
  {author} {\bibfnamefont {M.}~\bibnamefont {Dickmann}},\ }\bibfield  {title}
  {\bibinfo {title} {Messung experimenteller {K}ompetenz – ein
  computergestützter {E}xperimentiertest.},\ }\href@noop {} {\bibfield
  {journal} {\bibinfo  {journal} {PhyDid A}\ }\textbf {\bibinfo {volume} {1}},\
  \bibinfo {pages} {26} (\bibinfo {year} {2016})}\BibitemShut {NoStop}%
\bibitem [{\citenamefont {Rehfeldt}\ and\ \citenamefont
  {Nordmeier}(2018)}]{Rehfeldt}%
  \BibitemOpen
  \bibfield  {author} {\bibinfo {author} {\bibfnamefont {D.}~\bibnamefont
  {Rehfeldt}}\ and\ \bibinfo {author} {\bibfnamefont {V.}~\bibnamefont
  {Nordmeier}},\ }\bibfield  {title} {\bibinfo {title} {Lehrqualität
  naturwissenschaftlicher {H}ochschulpraktika. befunde zu {C}hemie- und
  {P}hysikpraktika, sowie {B}lock- und {S}emesterpraktika.},\ }\href@noop {}
  {\bibfield  {journal} {\bibinfo  {journal} {PhyDid}\ }\textbf {\bibinfo
  {volume} {1}},\ \bibinfo {pages} {34} (\bibinfo {year} {2018})}\BibitemShut
  {NoStop}%
\bibitem [{\citenamefont {Wilcox}\ and\ \citenamefont
  {Lewandowski}(2016{\natexlab{a}})}]{Wilcox16}%
  \BibitemOpen
  \bibfield  {author} {\bibinfo {author} {\bibfnamefont {B.~R.}\ \bibnamefont
  {Wilcox}}\ and\ \bibinfo {author} {\bibfnamefont {H.~J.}\ \bibnamefont
  {Lewandowski}},\ }\bibfield  {title} {\bibinfo {title} {Students’
  epistemologies about experimental physics: Validating the {C}olorado
  {L}earning {A}ttitudes about {S}cience {S}urvey for {E}xperimental
  {P}hysics},\ }\href@noop {} {\bibfield  {journal} {\bibinfo  {journal} {Phys.
  Rev. Phys. Educ. Res.}\ }\textbf {\bibinfo {volume} {12}},\ \bibinfo {pages}
  {010123} (\bibinfo {year} {2016}{\natexlab{a}})}\BibitemShut {NoStop}%
\bibitem [{\citenamefont {Priemer}(2003)}]{Priemer}%
  \BibitemOpen
  \bibfield  {author} {\bibinfo {author} {\bibfnamefont {B.}~\bibnamefont
  {Priemer}},\ }\bibfield  {title} {\bibinfo {title} {Ein diagnostischer {T}est
  zu {S}chüleransichten über {P}hysik und {L}ernen von {P}hysik – eine
  deutsche {V}ersion des {T}ests {V}iews {A}bout {S}cience {S}urvey},\
  }\href@noop {} {\bibfield  {journal} {\bibinfo  {journal} {Zeitschrift für
  Didaktik der Naturwissenschaften}\ }\textbf {\bibinfo {volume} {9}},\
  \bibinfo {pages} {160} (\bibinfo {year} {2003})}\BibitemShut {NoStop}%
\bibitem [{\citenamefont {Michel}\ and\ \citenamefont
  {Neumann}(2016)}]{Neumann}%
  \BibitemOpen
  \bibfield  {author} {\bibinfo {author} {\bibfnamefont {H.}~\bibnamefont
  {Michel}}\ and\ \bibinfo {author} {\bibfnamefont {I.}~\bibnamefont
  {Neumann}},\ }\bibfield  {title} {\bibinfo {title} {Nature of {S}cience and
  {S}cience {C}ontent {L}earning {T}he {R}elation {B}etween {S}tudents’
  {N}ature of {S}cience {U}nderstanding and {T}heir {L}earning {A}bout the
  {C}oncept of {E}nergy},\ }\href {https://doi.org/DOI
  10.1007/s11191-016-9860-4} {\bibfield  {journal} {\bibinfo  {journal} {Sci.
  Educ.}\ }\textbf {\bibinfo {volume} {25}},\ \bibinfo {pages} {951} (\bibinfo
  {year} {2016})}\BibitemShut {NoStop}%
\bibitem [{\citenamefont {Woitkowski}\ \emph {et~al.}(2021)\citenamefont
  {Woitkowski}, \citenamefont {Rochell},\ and\ \citenamefont {Bauer}}]{Woit21}%
  \BibitemOpen
  \bibfield  {author} {\bibinfo {author} {\bibfnamefont {D.}~\bibnamefont
  {Woitkowski}}, \bibinfo {author} {\bibfnamefont {L.}~\bibnamefont
  {Rochell}},\ and\ \bibinfo {author} {\bibfnamefont {A.~B.}\ \bibnamefont
  {Bauer}},\ }\bibfield  {title} {\bibinfo {title} {German university students'
  views of nature of science in the introductory phase},\ }\href
  {https://doi.org/10.1103/PhysRevPhysEducRes.17.010118} {\bibfield  {journal}
  {\bibinfo  {journal} {Phys. Rev. Phys. Educ. Res.}\ }\textbf {\bibinfo
  {volume} {17}},\ \bibinfo {pages} {010118} (\bibinfo {year}
  {2021})}\BibitemShut {NoStop}%
\bibitem [{\citenamefont {Wilcox}\ \emph {et~al.}(2016)\citenamefont {Wilcox},
  \citenamefont {Zwickl}, \citenamefont {Hobbs}, \citenamefont {Aiken},
  \citenamefont {Welch},\ and\ \citenamefont {Lewandowski}}]{WilcoxZwi16}%
  \BibitemOpen
  \bibfield  {author} {\bibinfo {author} {\bibfnamefont {B.~R.}\ \bibnamefont
  {Wilcox}}, \bibinfo {author} {\bibfnamefont {B.~M.}\ \bibnamefont {Zwickl}},
  \bibinfo {author} {\bibfnamefont {R.~D.}\ \bibnamefont {Hobbs}}, \bibinfo
  {author} {\bibfnamefont {J.~M.}\ \bibnamefont {Aiken}}, \bibinfo {author}
  {\bibfnamefont {N.~M.}\ \bibnamefont {Welch}},\ and\ \bibinfo {author}
  {\bibfnamefont {H.~J.}\ \bibnamefont {Lewandowski}},\ }\bibfield  {title}
  {\bibinfo {title} {Alternative model for administration and analysis of
  research-based assessments},\ }\href@noop {} {\bibfield  {journal} {\bibinfo
  {journal} {Phys. Rev. Phys. Educ. Res.}\ }\textbf {\bibinfo {volume} {12}},\
  \bibinfo {pages} {010139} (\bibinfo {year} {2016})}\BibitemShut {NoStop}%
\bibitem [{\citenamefont {Wilcox}\ and\ \citenamefont
  {Lewandowski}(2016{\natexlab{b}})}]{WilcoxLew16}%
  \BibitemOpen
  \bibfield  {author} {\bibinfo {author} {\bibfnamefont {B.~R.}\ \bibnamefont
  {Wilcox}}\ and\ \bibinfo {author} {\bibfnamefont {H.~J.}\ \bibnamefont
  {Lewandowski}},\ }\bibfield  {title} {\bibinfo {title} {Open-ended versus
  guided laboratory activities: Impact on students’ beliefs about
  experimental physics},\ }\href@noop {} {\bibfield  {journal} {\bibinfo
  {journal} {Phys. Rev. Phys. Educ. Res.}\ }\textbf {\bibinfo {volume} {12}},\
  \bibinfo {pages} {020132} (\bibinfo {year} {2016}{\natexlab{b}})}\BibitemShut
  {NoStop}%
\bibitem [{\citenamefont {Wilcox}\ and\ \citenamefont
  {Lewandowski}(2017{\natexlab{a}})}]{WilcoxLew17}%
  \BibitemOpen
  \bibfield  {author} {\bibinfo {author} {\bibfnamefont {B.~R.}\ \bibnamefont
  {Wilcox}}\ and\ \bibinfo {author} {\bibfnamefont {H.~J.}\ \bibnamefont
  {Lewandowski}},\ }\bibfield  {title} {\bibinfo {title} {Developing skills
  versus reinforcing concepts in physics labs: Insight from a survey of
  students’ beliefs about experimental physics},\ }\href@noop {} {\bibfield
  {journal} {\bibinfo  {journal} {Phys. Rev. Phys. Educ. Res.}\ }\textbf
  {\bibinfo {volume} {13}},\ \bibinfo {pages} {010108} (\bibinfo {year}
  {2017}{\natexlab{a}})}\BibitemShut {NoStop}%
\bibitem [{\citenamefont {Fox}\ \emph {et~al.}(2021{\natexlab{a}})\citenamefont
  {Fox}, \citenamefont {Hoehn}, \citenamefont {Werth},\ and\ \citenamefont
  {Lewandowski}}]{MichaelJFox}%
  \BibitemOpen
  \bibfield  {author} {\bibinfo {author} {\bibfnamefont {M.~F.~J.}\
  \bibnamefont {Fox}}, \bibinfo {author} {\bibfnamefont {J.~R.}\ \bibnamefont
  {Hoehn}}, \bibinfo {author} {\bibfnamefont {A.}~\bibnamefont {Werth}},\ and\
  \bibinfo {author} {\bibfnamefont {H.~J.}\ \bibnamefont {Lewandowski}},\
  }\bibfield  {title} {\bibinfo {title} {Lab instruction during the covid-19
  pandemic: Effects on student views about experimental physics in comparison
  with previous years},\ }\href@noop {} {\bibfield  {journal} {\bibinfo
  {journal} {Phys. Rev. Phys. Educ. Res.}\ }\textbf {\bibinfo {volume} {17}},\
  \bibinfo {pages} {010148} (\bibinfo {year} {2021}{\natexlab{a}})}\BibitemShut
  {NoStop}%
\bibitem [{\citenamefont {Aiken}\ and\ \citenamefont
  {Lewandowski}(2021)}]{Aiken&Lew}%
  \BibitemOpen
  \bibfield  {author} {\bibinfo {author} {\bibfnamefont {J.~M.}\ \bibnamefont
  {Aiken}}\ and\ \bibinfo {author} {\bibfnamefont {H.~J.}\ \bibnamefont
  {Lewandowski}},\ }\bibfield  {title} {\bibinfo {title} {Data sharing model
  for physics education research using the 70000 response colorado learning
  attitudes about science survey for experimental physics data set},\
  }\href@noop {} {\bibfield  {journal} {\bibinfo  {journal} {Phys. Rev. Phys.
  Educ. Res.}\ }\textbf {\bibinfo {volume} {17}},\ \bibinfo {pages} {020144}
  (\bibinfo {year} {2021})}\BibitemShut {NoStop}%
\bibitem [{\citenamefont {Wilcox}\ and\ \citenamefont
  {Lewandowski}(2017{\natexlab{b}})}]{WilcoxLew17PERC}%
  \BibitemOpen
  \bibfield  {author} {\bibinfo {author} {\bibfnamefont {B.~R.}\ \bibnamefont
  {Wilcox}}\ and\ \bibinfo {author} {\bibfnamefont {H.~J.}\ \bibnamefont
  {Lewandowski}},\ }\bibfield  {title} {\bibinfo {title} {Impact of perceived
  grading practices on students’ beliefs about experimental physics},\
  }\href@noop {} {\bibfield  {journal} {\bibinfo  {journal} {PER Conf. Proc.}\
  } (\bibinfo {year} {2017}{\natexlab{b}})}\BibitemShut {NoStop}%
\bibitem [{\citenamefont {Van~Dusen}\ \emph {et~al.}(2021)\citenamefont
  {Van~Dusen}, \citenamefont {Shultz}, \citenamefont {Nissen}, \citenamefont
  {Wilcox}, \citenamefont {Holmes}, \citenamefont {Jariwala}, \citenamefont
  {Close}, \citenamefont {Lewandowski},\ and\ \citenamefont
  {Pollock}}]{vandusen}%
  \BibitemOpen
  \bibfield  {author} {\bibinfo {author} {\bibfnamefont {B.}~\bibnamefont
  {Van~Dusen}}, \bibinfo {author} {\bibfnamefont {M.}~\bibnamefont {Shultz}},
  \bibinfo {author} {\bibfnamefont {J.~M.}\ \bibnamefont {Nissen}}, \bibinfo
  {author} {\bibfnamefont {B.~R.}\ \bibnamefont {Wilcox}}, \bibinfo {author}
  {\bibfnamefont {N.~G.}\ \bibnamefont {Holmes}}, \bibinfo {author}
  {\bibfnamefont {M.}~\bibnamefont {Jariwala}}, \bibinfo {author}
  {\bibfnamefont {E.~W.}\ \bibnamefont {Close}}, \bibinfo {author}
  {\bibfnamefont {H.~J.}\ \bibnamefont {Lewandowski}},\ and\ \bibinfo {author}
  {\bibfnamefont {S.}~\bibnamefont {Pollock}},\ }\bibfield  {title} {\bibinfo
  {title} {Online administration of research-based assessments},\ }\href
  {https://doi.org/10.1119/10.0002888} {\bibfield  {journal} {\bibinfo
  {journal} {American Journal of Physics}\ }\textbf {\bibinfo {volume} {89}},\
  \bibinfo {pages} {7} (\bibinfo {year} {2021})},\ \Eprint
  {https://arxiv.org/abs/https://doi.org/10.1119/10.0002888}
  {https://doi.org/10.1119/10.0002888} \BibitemShut {NoStop}%
\bibitem [{Web()}]{WebpageGE-CLASS}%
  \BibitemOpen
  \href@noop {} {\bibinfo {title}
  {\url{http://geclass.physik.uni-potsdam.de/auth/login}}}\BibitemShut
  {NoStop}%
\bibitem [{\citenamefont {Van~Rossum}\ and\ \citenamefont
  {Drake}(2009)}]{10.5555/1593511}%
  \BibitemOpen
  \bibfield  {author} {\bibinfo {author} {\bibfnamefont {G.}~\bibnamefont
  {Van~Rossum}}\ and\ \bibinfo {author} {\bibfnamefont {F.~L.}\ \bibnamefont
  {Drake}},\ }\href@noop {} {\emph {\bibinfo {title} {Python 3 Reference
  Manual}}}\ (\bibinfo  {publisher} {CreateSpace},\ \bibinfo {address} {Scotts
  Valley, CA},\ \bibinfo {year} {2009})\BibitemShut {NoStop}%
\bibitem [{\citenamefont {Virtanen}\ \emph {et~al.}(2020)\citenamefont
  {Virtanen}, \citenamefont {Gommers}, \citenamefont {Oliphant}, \citenamefont
  {Haberland}, \citenamefont {Reddy}, \citenamefont {Cournapeau}, \citenamefont
  {Burovski}, \citenamefont {Peterson}, \citenamefont {Weckesser},
  \citenamefont {Bright}, \citenamefont {{van der Walt}}, \citenamefont
  {Brett}, \citenamefont {Wilson}, \citenamefont {Millman}, \citenamefont
  {Mayorov}, \citenamefont {Nelson}, \citenamefont {Jones}, \citenamefont
  {Kern}, \citenamefont {Larson}, \citenamefont {Carey}, \citenamefont {Polat},
  \citenamefont {Feng}, \citenamefont {Moore}, \citenamefont {{VanderPlas}},
  \citenamefont {Laxalde}, \citenamefont {Perktold}, \citenamefont {Cimrman},
  \citenamefont {Henriksen}, \citenamefont {Quintero}, \citenamefont {Harris},
  \citenamefont {Archibald}, \citenamefont {Ribeiro}, \citenamefont
  {Pedregosa}, \citenamefont {{van Mulbregt}},\ and\ \citenamefont {{SciPy 1.0
  Contributors}}}]{2020SciPy-NMeth}%
  \BibitemOpen
  \bibfield  {author} {\bibinfo {author} {\bibfnamefont {P.}~\bibnamefont
  {Virtanen}}, \bibinfo {author} {\bibfnamefont {R.}~\bibnamefont {Gommers}},
  \bibinfo {author} {\bibfnamefont {T.~E.}\ \bibnamefont {Oliphant}}, \bibinfo
  {author} {\bibfnamefont {M.}~\bibnamefont {Haberland}}, \bibinfo {author}
  {\bibfnamefont {T.}~\bibnamefont {Reddy}}, \bibinfo {author} {\bibfnamefont
  {D.}~\bibnamefont {Cournapeau}}, \bibinfo {author} {\bibfnamefont
  {E.}~\bibnamefont {Burovski}}, \bibinfo {author} {\bibfnamefont
  {P.}~\bibnamefont {Peterson}}, \bibinfo {author} {\bibfnamefont
  {W.}~\bibnamefont {Weckesser}}, \bibinfo {author} {\bibfnamefont
  {J.}~\bibnamefont {Bright}}, \bibinfo {author} {\bibfnamefont {S.~J.}\
  \bibnamefont {{van der Walt}}}, \bibinfo {author} {\bibfnamefont
  {M.}~\bibnamefont {Brett}}, \bibinfo {author} {\bibfnamefont
  {J.}~\bibnamefont {Wilson}}, \bibinfo {author} {\bibfnamefont {K.~J.}\
  \bibnamefont {Millman}}, \bibinfo {author} {\bibfnamefont {N.}~\bibnamefont
  {Mayorov}}, \bibinfo {author} {\bibfnamefont {A.~R.~J.}\ \bibnamefont
  {Nelson}}, \bibinfo {author} {\bibfnamefont {E.}~\bibnamefont {Jones}},
  \bibinfo {author} {\bibfnamefont {R.}~\bibnamefont {Kern}}, \bibinfo {author}
  {\bibfnamefont {E.}~\bibnamefont {Larson}}, \bibinfo {author} {\bibfnamefont
  {C.~J.}\ \bibnamefont {Carey}}, \bibinfo {author} {\bibfnamefont
  {{\.I}.}~\bibnamefont {Polat}}, \bibinfo {author} {\bibfnamefont
  {Y.}~\bibnamefont {Feng}}, \bibinfo {author} {\bibfnamefont {E.~W.}\
  \bibnamefont {Moore}}, \bibinfo {author} {\bibfnamefont {J.}~\bibnamefont
  {{VanderPlas}}}, \bibinfo {author} {\bibfnamefont {D.}~\bibnamefont
  {Laxalde}}, \bibinfo {author} {\bibfnamefont {J.}~\bibnamefont {Perktold}},
  \bibinfo {author} {\bibfnamefont {R.}~\bibnamefont {Cimrman}}, \bibinfo
  {author} {\bibfnamefont {I.}~\bibnamefont {Henriksen}}, \bibinfo {author}
  {\bibfnamefont {E.~A.}\ \bibnamefont {Quintero}}, \bibinfo {author}
  {\bibfnamefont {C.~R.}\ \bibnamefont {Harris}}, \bibinfo {author}
  {\bibfnamefont {A.~M.}\ \bibnamefont {Archibald}}, \bibinfo {author}
  {\bibfnamefont {A.~H.}\ \bibnamefont {Ribeiro}}, \bibinfo {author}
  {\bibfnamefont {F.}~\bibnamefont {Pedregosa}}, \bibinfo {author}
  {\bibfnamefont {P.}~\bibnamefont {{van Mulbregt}}},\ and\ \bibinfo {author}
  {\bibnamefont {{SciPy 1.0 Contributors}}},\ }\bibfield  {title} {\bibinfo
  {title} {{{SciPy} 1.0: Fundamental Algorithms for Scientific Computing in
  Python}},\ }\href {https://doi.org/10.1038/s41592-019-0686-2} {\bibfield
  {journal} {\bibinfo  {journal} {Nature Methods}\ }\textbf {\bibinfo {volume}
  {17}},\ \bibinfo {pages} {261} (\bibinfo {year} {2020})}\BibitemShut
  {NoStop}%
\bibitem [{\citenamefont {Harris}\ \emph {et~al.}(2020)\citenamefont {Harris},
  \citenamefont {Millman}, \citenamefont {van~der Walt}, \citenamefont
  {Gommers}, \citenamefont {Virtanen}, \citenamefont {Cournapeau},
  \citenamefont {Wieser}, \citenamefont {Taylor}, \citenamefont {Berg},
  \citenamefont {Smith}, \citenamefont {Kern}, \citenamefont {Picus},
  \citenamefont {Hoyer}, \citenamefont {van Kerkwijk}, \citenamefont {Brett},
  \citenamefont {Haldane}, \citenamefont {Fernández~del Río}, \citenamefont
  {Wiebe}, \citenamefont {Peterson}, \citenamefont {Gérard-Marchant},
  \citenamefont {Sheppard}, \citenamefont {Reddy}, \citenamefont {Weckesser},
  \citenamefont {Abbasi}, \citenamefont {Gohlke},\ and\ \citenamefont
  {Oliphant}}]{2020NumPy-Array}%
  \BibitemOpen
  \bibfield  {author} {\bibinfo {author} {\bibfnamefont {C.~R.}\ \bibnamefont
  {Harris}}, \bibinfo {author} {\bibfnamefont {K.~J.}\ \bibnamefont {Millman}},
  \bibinfo {author} {\bibfnamefont {S.~J.}\ \bibnamefont {van~der Walt}},
  \bibinfo {author} {\bibfnamefont {R.}~\bibnamefont {Gommers}}, \bibinfo
  {author} {\bibfnamefont {P.}~\bibnamefont {Virtanen}}, \bibinfo {author}
  {\bibfnamefont {D.}~\bibnamefont {Cournapeau}}, \bibinfo {author}
  {\bibfnamefont {E.}~\bibnamefont {Wieser}}, \bibinfo {author} {\bibfnamefont
  {J.}~\bibnamefont {Taylor}}, \bibinfo {author} {\bibfnamefont
  {S.}~\bibnamefont {Berg}}, \bibinfo {author} {\bibfnamefont {N.~J.}\
  \bibnamefont {Smith}}, \bibinfo {author} {\bibfnamefont {R.}~\bibnamefont
  {Kern}}, \bibinfo {author} {\bibfnamefont {M.}~\bibnamefont {Picus}},
  \bibinfo {author} {\bibfnamefont {S.}~\bibnamefont {Hoyer}}, \bibinfo
  {author} {\bibfnamefont {M.~H.}\ \bibnamefont {van Kerkwijk}}, \bibinfo
  {author} {\bibfnamefont {M.}~\bibnamefont {Brett}}, \bibinfo {author}
  {\bibfnamefont {A.}~\bibnamefont {Haldane}}, \bibinfo {author} {\bibfnamefont
  {J.}~\bibnamefont {Fernández~del Río}}, \bibinfo {author} {\bibfnamefont
  {M.}~\bibnamefont {Wiebe}}, \bibinfo {author} {\bibfnamefont
  {P.}~\bibnamefont {Peterson}}, \bibinfo {author} {\bibfnamefont
  {P.}~\bibnamefont {Gérard-Marchant}}, \bibinfo {author} {\bibfnamefont
  {K.}~\bibnamefont {Sheppard}}, \bibinfo {author} {\bibfnamefont
  {T.}~\bibnamefont {Reddy}}, \bibinfo {author} {\bibfnamefont
  {W.}~\bibnamefont {Weckesser}}, \bibinfo {author} {\bibfnamefont
  {H.}~\bibnamefont {Abbasi}}, \bibinfo {author} {\bibfnamefont
  {C.}~\bibnamefont {Gohlke}},\ and\ \bibinfo {author} {\bibfnamefont {T.~E.}\
  \bibnamefont {Oliphant}},\ }\bibfield  {title} {\bibinfo {title} {Array
  programming with {NumPy}},\ }\href
  {https://doi.org/10.1038/s41586-020-2649-2} {\bibfield  {journal} {\bibinfo
  {journal} {Nature}\ }\textbf {\bibinfo {volume} {585}},\ \bibinfo {pages}
  {357–362} (\bibinfo {year} {2020})}\BibitemShut {NoStop}%
\bibitem [{\citenamefont {{Hunter}}(2007)}]{4160265}%
  \BibitemOpen
  \bibfield  {author} {\bibinfo {author} {\bibfnamefont {J.~D.}\ \bibnamefont
  {{Hunter}}},\ }\bibfield  {title} {\bibinfo {title} {Matplotlib: A 2d
  graphics environment},\ }\href {https://doi.org/10.1109/MCSE.2007.55}
  {\bibfield  {journal} {\bibinfo  {journal} {Computing in Science
  Engineering}\ }\textbf {\bibinfo {volume} {9}},\ \bibinfo {pages} {90}
  (\bibinfo {year} {2007})}\BibitemShut {NoStop}%
\bibitem [{rep()}]{report}%
  \BibitemOpen
  \href@noop {} {\bibinfo {title}
  {\url{http://geclass.physik.uni-potsdam.de/report/example.pdf}}}\BibitemShut
  {NoStop}%
\bibitem [{\citenamefont {Dounas-Frazer}\ and\ \citenamefont
  {Lewandowski}(2018)}]{Dounas_Frazer_2018}%
  \BibitemOpen
  \bibfield  {author} {\bibinfo {author} {\bibfnamefont {D.~R.}\ \bibnamefont
  {Dounas-Frazer}}\ and\ \bibinfo {author} {\bibfnamefont {H.~J.}\ \bibnamefont
  {Lewandowski}},\ }\bibfield  {title} {\bibinfo {title} {The modelling
  framework for experimental physics: description, development, and
  applications},\ }\href {https://doi.org/10.1088/1361-6404/aae3ce} {\bibfield
  {journal} {\bibinfo  {journal} {European Journal of Physics}\ }\textbf
  {\bibinfo {volume} {39}},\ \bibinfo {pages} {064005} (\bibinfo {year}
  {2018})}\BibitemShut {NoStop}%
\bibitem [{\citenamefont {Mann}\ and\ \citenamefont
  {Whitney}(1947)}]{MannWhitney47}%
  \BibitemOpen
  \bibfield  {author} {\bibinfo {author} {\bibfnamefont {H.~B.}\ \bibnamefont
  {Mann}}\ and\ \bibinfo {author} {\bibfnamefont {D.~R.}\ \bibnamefont
  {Whitney}},\ }\bibfield  {title} {\bibinfo {title} {On a test of whether one
  of two random variables is stochastically larger than the other},\
  }\href@noop {} {\bibfield  {journal} {\bibinfo  {journal} {Ann. Math. Stat.}\
  }\textbf {\bibinfo {volume} {18}},\ \bibinfo {pages} {50} (\bibinfo {year}
  {1947})}\BibitemShut {NoStop}%
\bibitem [{\citenamefont {Scholz}\ and\ \citenamefont
  {Stephens}(1987)}]{K-SampleAnderson}%
  \BibitemOpen
  \bibfield  {author} {\bibinfo {author} {\bibfnamefont {F.~W.}\ \bibnamefont
  {Scholz}}\ and\ \bibinfo {author} {\bibfnamefont {M.~A.}\ \bibnamefont
  {Stephens}},\ }\bibfield  {title} {\bibinfo {title} {K-sample
  anderson-darling tests},\ }\href@noop {} {\bibfield  {journal} {\bibinfo
  {journal} {Journal of the American Statistical Association}\ }\textbf
  {\bibinfo {volume} {82}},\ \bibinfo {pages} {918} (\bibinfo {year}
  {1987})}\BibitemShut {NoStop}%
\bibitem [{\citenamefont {Cohen}(1988)}]{Cohen88}%
  \BibitemOpen
  \bibfield  {author} {\bibinfo {author} {\bibfnamefont {J.}~\bibnamefont
  {Cohen}},\ }\href@noop {} {\emph {\bibinfo {title} {Statistical Power
  Analysis for the Behavioral Sciences}}}\ (\bibinfo  {publisher} {Lawrence
  Erlbaum Associates, Hillsdale, NJ},\ \bibinfo {year} {1988})\BibitemShut
  {NoStop}%
\bibitem [{\citenamefont {Gray}\ \emph {et~al.}(2008)\citenamefont {Gray},
  \citenamefont {Adams}, \citenamefont {Wieman},\ and\ \citenamefont
  {Perkins}}]{Gray08}%
  \BibitemOpen
  \bibfield  {author} {\bibinfo {author} {\bibfnamefont {K.~E.}\ \bibnamefont
  {Gray}}, \bibinfo {author} {\bibfnamefont {W.~K.}\ \bibnamefont {Adams}},
  \bibinfo {author} {\bibfnamefont {C.~E.}\ \bibnamefont {Wieman}},\ and\
  \bibinfo {author} {\bibfnamefont {K.~K.}\ \bibnamefont {Perkins}},\
  }\bibfield  {title} {\bibinfo {title} {Students know what physicists believe,
  but they don’t agree: A study using the class survey},\ }\href@noop {}
  {\bibfield  {journal} {\bibinfo  {journal} {Phys. Rev. ST Phys. Educ. Res.}\
  }\textbf {\bibinfo {volume} {0}},\ \bibinfo {pages} {020106} (\bibinfo {year}
  {2008})}\BibitemShut {NoStop}%
\bibitem [{\citenamefont {S.~Allie}\ and\ \citenamefont
  {Lubben}(1998)}]{Allie}%
  \BibitemOpen
  \bibfield  {author} {\bibinfo {author} {\bibfnamefont {B.~C.}\ \bibnamefont
  {S.~Allie}, \bibfnamefont {A.~Buffler}}\ and\ \bibinfo {author}
  {\bibfnamefont {F.}~\bibnamefont {Lubben}},\ }\bibfield  {title} {\bibinfo
  {title} {First-year physics students’ perceptions of the quality of
  experimental measurements},\ }\href@noop {} {\bibfield  {journal} {\bibinfo
  {journal} {International Journal of Science Education}\ }\textbf {\bibinfo
  {volume} {20}},\ \bibinfo {pages} {447} (\bibinfo {year} {1998})}\BibitemShut
  {NoStop}%
\bibitem [{\citenamefont {Pollard}\ \emph {et~al.}(2017)\citenamefont
  {Pollard}, \citenamefont {Hobbs}, \citenamefont {Stanley}, \citenamefont
  {Dounas-Frazer},\ and\ \citenamefont {Lewandowski}}]{Pollard2017}%
  \BibitemOpen
  \bibfield  {author} {\bibinfo {author} {\bibfnamefont {B.}~\bibnamefont
  {Pollard}}, \bibinfo {author} {\bibfnamefont {R.}~\bibnamefont {Hobbs}},
  \bibinfo {author} {\bibfnamefont {J.~T.}\ \bibnamefont {Stanley}}, \bibinfo
  {author} {\bibfnamefont {D.}~\bibnamefont {Dounas-Frazer}},\ and\ \bibinfo
  {author} {\bibfnamefont {H.~J.}\ \bibnamefont {Lewandowski}},\ }\bibfield
  {title} {\bibinfo {title} {Impact of an introductory lab course on
  students’ understanding of measurement uncertainty},\ }in\ \href@noop {}
  {\emph {\bibinfo {booktitle} {Physics Education Research Conference 2017}}},\
  \bibinfo {series and number} {PER Conference}\ (\bibinfo {address}
  {Cincinnati, OH},\ \bibinfo {year} {2017})\ pp.\ \bibinfo {pages}
  {312--315}\BibitemShut {NoStop}%
\bibitem [{\citenamefont {Pollard}\ \emph {et~al.}(2020)\citenamefont
  {Pollard}, \citenamefont {Werth}, \citenamefont {Hobbs},\ and\ \citenamefont
  {Lewandowski}}]{Pollard2020}%
  \BibitemOpen
  \bibfield  {author} {\bibinfo {author} {\bibfnamefont {B.}~\bibnamefont
  {Pollard}}, \bibinfo {author} {\bibfnamefont {A.}~\bibnamefont {Werth}},
  \bibinfo {author} {\bibfnamefont {R.}~\bibnamefont {Hobbs}},\ and\ \bibinfo
  {author} {\bibfnamefont {H.~J.}\ \bibnamefont {Lewandowski}},\ }\bibfield
  {title} {\bibinfo {title} {Impact of a course transformation on students’
  reasoning about measurement uncertainty},\ }\href@noop {} {\bibfield
  {journal} {\bibinfo  {journal} {Phys. Rev. Phys. Educ. Res.}\ }\textbf
  {\bibinfo {volume} {16}},\ \bibinfo {pages} {020160} (\bibinfo {year}
  {2020})}\BibitemShut {NoStop}%
\bibitem [{\citenamefont {Fox}\ \emph {et~al.}(2021{\natexlab{b}})\citenamefont
  {Fox}, \citenamefont {Hoehn}, \citenamefont {Werth},\ and\ \citenamefont
  {Lewandowski}}]{FoxEclasssPandemic}%
  \BibitemOpen
  \bibfield  {author} {\bibinfo {author} {\bibfnamefont {M.~F.~J.}\
  \bibnamefont {Fox}}, \bibinfo {author} {\bibfnamefont {J.~R.}\ \bibnamefont
  {Hoehn}}, \bibinfo {author} {\bibfnamefont {A.}~\bibnamefont {Werth}},\ and\
  \bibinfo {author} {\bibfnamefont {H.~J.}\ \bibnamefont {Lewandowski}},\
  }\bibfield  {title} {\bibinfo {title} {Lab instruction during the covid-19
  pandemic: Effects on student views about experimental physics in comparison
  with previous years},\ }\href
  {https://doi.org/10.1103/PhysRevPhysEducRes.17.010148} {\bibfield  {journal}
  {\bibinfo  {journal} {Phys. Rev. Phys. Educ. Res.}\ }\textbf {\bibinfo
  {volume} {17}},\ \bibinfo {pages} {010148} (\bibinfo {year}
  {2021}{\natexlab{b}})}\BibitemShut {NoStop}%
\end{thebibliography}
\end{document}